\documentclass[lettersize,journal]{IEEEtran}
\IEEEoverridecommandlockouts
\usepackage[hyphens]{url}
\usepackage{breakurl}
\usepackage{cite}
\usepackage{citesort}
\usepackage{amsmath,amsfonts,amsthm,amssymb}
\usepackage{bbm}
\usepackage{algpseudocode}
\usepackage{algorithm}

\usepackage{graphicx}
\usepackage{subfigure}
\graphicspath{{figure/}}
\usepackage{textcomp}
\usepackage{booktabs}
\usepackage{setspace}
\usepackage{tabularx}
\newcommand{\ra}[1]{\renewcommand{\arraystretch}{#1}}

\newcolumntype{L}{>{\hspace*{-\tabcolsep}}l}
\newcolumntype{R}{c<{\hspace*{-\tabcolsep}}}
\usepackage[table]{xcolor}
\usepackage{xcolor}
\usepackage{extarrows}
\usepackage{times}
\usepackage[justification = centering]{caption}
\definecolor{lightblue}{rgb}{0.93,0.95,1.0}

\def\BibTeX{{\rm B\kern-.05em{\sc i\kern-.025em b}\kern-.08em
		T\kern-.1667em\lower.7ex\hbox{E}\kern-.125emX}}

\newcommand{\cE}{\mathcal{E}}

\newcommand{\cG}{\mathcal{G}}
\newcommand{\cH}{\mathcal{H}}

\newcommand{\cK}{\mathcal{K}}
\newcommand{\cL}{\mathcal{L}}

\newcommand{\CN}{\mathcal{CN}}
\newcommand{\cO}{\mathcal{O}}

\newcommand{\cS}{\mathcal{S}}

\newcommand{\cV}{\mathcal{V}}

\newcommand{\be}{\mathbf{e}}

\newcommand{\bh}{\mathbf{h}}

\newcommand{\br}{\mathbf{r}}
\newcommand{\bs}{\mathbf{s}}

\newcommand{\bu}{\mathbf{u}}

\newcommand{\bx}{\mathbf{x}}

\newcommand{\bz}{\mathbf{z}}

\newcommand{\bG}{\mathbf{G}}
\newcommand{\bH}{\mathbf{H}}
\newcommand{\bI}{\mathbf{I}}
\newcommand{\bJ}{\mathbf{J}}

\newcommand{\bP}{\mathbf{P}}
\newcommand{\bQ}{\mathbf{Q}}

\newcommand{\bX}{\mathbf{X}}

\newcommand{\rT}{\mathrm{T}}

\newcommand{\bbC}{\mathbb{C}}
\newcommand{\bbR}{\mathbb{R}}

\newcommand{\bzero}{\mathbf{0}}

\newcommand{\bbeta}{\boldsymbol\beta}

\newcommand{\bmu}{\boldsymbol\mu}

\newcommand{\blambda}{{\boldsymbol\lambda}}

\newcommand{\brho}{{\boldsymbol\rho}}

\newcommand{\figref}[1]{Fig.~\ref{#1}}
\newcommand{\secref}[1]{Section~\ref{#1}}
\newcommand{\subsecref}[1]{Subsection~\ref{#1}}
\newcommand{\alref}[1]{\textbf{Algorithm}~\textbf{\ref{#1}}}

\newcommand{\lemmref}[1]{\textit{Lemma}~\ref{#1}}
\newcommand{\theoref}[1]{\textit{Theorem}~\ref{#1}}
\newcommand{\propref}[1]{\textit{Proposition}~\ref{#1}}
\newcommand{\tabref}[1]{Table~\ref{#1}}

\newcommand{\st}{\mathrm{s.t.}}

\newcommand{\rdiag}[1]{\mathrm{diag}\left\{ #1\right\}}
\newcommand{\trace}[1]{\mathrm{tr}\left(#1\right)}
\newcommand{\expect}[1]{\mathbb{E}{\left\{#1\right\}}}

\newcommand{\real}[1]{\Re\left\{#1\right\}}

\newtheorem{theorem}{Theorem}

\newtheorem{lemma}{Lemma}

\newtheorem{proposition}{Proposition}

\begin{document} 
\captionsetup{justification=raggedright,singlelinecheck=false}
\title{Continuous Aperture Array (CAPA)-Based Multi-Group Multicast Communications} 
\author{
	Mengyu~Qian, \IEEEmembership{Graduate Student Member,~IEEE,} 
	Xidong~Mu,~\IEEEmembership{Member,~IEEE,} 
	Li~You,~\IEEEmembership{Senior Member,~IEEE,}
    and Michail~Matthaiou,~\IEEEmembership{Fellow,~IEEE}
	\thanks{Mengyu Qian and Li You are with the National Mobile Communications Research Laboratory, Southeast University, Nanjing 210096, China, and also with the Purple Mountain Laboratories, Nanjing 211100, China (e-mail: \{qianmy, lyou\}@seu.edu.cn).}
	\thanks{Xidong Mu is with the Centre for Wireless Innovation (CWI), Queen's University Belfast, BT3 9DT Belfast, U.K. (e-mail:  x.mu@qub.ac.uk).}
\thanks{Michail Matthaiou is with the Centre for Wireless Innovation (CWI), Queen's University Belfast, BT3 9DT Belfast, U.K., and also affiliated with the Department of Electronic Engineering, Kyung Hee University, Yongin-si, Gyeonggi-do 17104, Republic of Korea (email: m.matthaiou@qub.ac.uk). The work of M. Matthaiou was supported by a research grant from the Department for the Economy Northern Ireland under the US-Ireland R\&D Partnership Programme and by the European Research Council (ERC) under the European Union's Horizon 2020 research and innovation programme (grant agreement No. 101001331).}
} 
\maketitle
\begin{abstract}
A continuous aperture array (CAPA)-based multi-group multicast communication system is investigated.  An integral-based CAPA multi-group multicast beamforming design is formulated for the maximization of the system energy efficiency (EE), subject to a minimum multicast SE constraint of each user group and a total transmit power constraint.  To address this non-econvex fractional programming problem, the Dinkelbach's method is employed. Within the Dinkelbach's framework,  the non-convex group-wise multicast spectral efficiency (SE) constraint is first equivalently transformed into a tractable form with auxiliary variables. Then, an efficient block coordinate descent (BCD)-based algorithm is developed to solve the reformulated problem.  The CAPA beamforming design subproblem can be optimally solved via the Lagrangian dual method and the calculus of variations (CoV) theory. It reveals that the optimal CAPA beamformer should be a combination of all the groups' user channels. To further reduce the computational complexity, a low-complexity zero-forcing (ZF)-based approach is proposed. The closed-form ZF CAPA beamformer is derived using each group's most representative user channel to mitigate the inter-group interference while ensuring the intra-group multicast performance. Then, the beamforming design subproblem in the BCD-based algorithm becomes a convex power allocation subproblem, which can be efficiently solved. Numerical results demonstrate that 1) the CAPA can significantly improve the EE compared to conventional spatially discrete arrays (SPDAs); 2) due to the enhanced spatial resolutions, increasing the aperture size of CAPA is not always beneficial for EE enhancement in multicast scenarios; and 3) wider user distributions of each group cause a significant EE degradation of CAPA compared to SPDA.
\end{abstract}


\section{Introduction}
During the rapid development of wireless communications, multiple-input multiple-output (MIMO) has become one of the most pioneering technologies \cite{10684260,zhang2020prospective,jin2024gdm4mmimo,10117500}. Compared to fifth-generation (5G), next-generation wireless networks, namely sixth generation (6G) and beyond (B6G), impose more stringent requirements, e.g., ultra-high data rate, massive connectivity, and highly accurate sensing. To meet these demands, one common approach is to further increase the number of antennas and the size of arrays, i.e., evolving MIMO to massive MIMO and extremely large-scale MIMO \cite{xuNearFieldWidebandExtremely2022a}. 
It is worth noting that these types of MIMO technologies rely on the employment of spatially discrete antenna arrays (SPDAs), where a large number of antennas are separately installed at a given aperture size. This, however, is inevitably constrained by the finite number of degrees-of-freedom (DoFs). Considering the fact that the connection demand of B6G is continuously growing without limitations, the relentless increase of MIMO dimensions leads to unaffordable hardware cost and energy consumption. Therefore, revolutionary technologies are needed.
 
In recent years, a new concept, known as continuous aperture array (CAPA),
has garnered significant research attention.
In contrast to SPDAs, CAPAs are fundamentally characterized by progressively reducing the antenna spacing to infinitely small within a given array aperture, in order to form a continuous EM radiating surface \cite{wei2024electromagnetic}.
Generally speaking, a CAPA can be regarded to have a spatially continuous architecture with a (virtually) infinite number of antennas. As a result, CAPAs are capable of manipulating the radiated signal's amplitudes and phase shifts of every single point on the surface through continuous source current patterns, a concept known as CAPA beamformer \cite{liu2024capa}. Compared to SPDAs, CAPAs provide significantly enhanced DoFs and array gains for the communication performance enhancement, thus becoming a promising technology for B6G \cite{huang2020holographic}.

\subsection{Prior Works} 
Given the unique characteristics of continuously manipulating signals, CAPAs have introduced fundamentally different signal and channel models compared to SPDAs. This also leads to new challenges pertaining to the performance analysis and performance optimization to characterize and reap the benefits of CAPA-based wireless communications. 
Motivated by the feature of having a virtually infinite number of antennas, some earlier research efforts have analyzed the DoFs achieved by CAPAs. 
For example, the authors of \cite{poonDegreesFreedomMultipleantenna2005} derived the available spatial DoFs of continuous spherical antenna arrays. Their work revealed that the spatial DoFs are related to the area and the geometry constraints of the array. Furthermore, the authors of \cite{poonDegreeFreedomGainUsing2011} provided a mathematical framework that can be applied to any array geometry and channel scattering conditions to analyze the DoFs.
In \cite{dardariCommunicatingLargeIntelligent2020}, the author derived analytical expressions for the link gain and the available spatial DoFs for large continuous intelligent surface-based line-of-sight (LoS) communications. In \cite{9848802}, the authors further analyzed the achievable DoFs for a linear continuous large-scale antenna array, which showed that the system DoFs are related to the spatial bandwidth.
\begin{table}[t]
\centering
\ra{1.3}
\footnotesize
\caption{Our contributions in contrast to the state-of-the-art}
\resizebox{\linewidth}{!}{
\begin{tabular}{LccccccR}
\toprule
&\hspace{-0.08cm}\cite{sanguinettiWavenumberDivisionMultiplexingLineofSight2023} 
&\hspace{-0.08cm}\cite{zhangPatternDivisionMultiplexingMultiUser2023} 
&\hspace{-0.08cm}\cite{qian2024spectral}
&\hspace{-0.08cm}\cite{wang2025beamforming}
&\hspace{-0.08cm}\cite{wang2025optimalbeamformingmultiusercontinuous}
&\hspace{-0.08cm}\cite{securebeamformingCAPA2025}   
&\hspace{-0.08cm}\textbf{Our Work} \\ 
\midrule \rowcolor{lightblue}
Unicast &\checkmark &\checkmark &\checkmark &\checkmark &\checkmark &\checkmark  &\text{Applicable} \\
Multi-group multicast &$\times$ &$\times$ &$\times$ &$\times$  &$\times$ &$\times$   &\checkmark \\ \rowcolor{lightblue}
Energy efficiency &$\times$  &$\times$  &$\times$ &$\times$ &$\times$ &$\times$ &\checkmark \\ 
Continuous beamformer & \multicolumn{3}{c}{\text{Discrete Approximation}} &\checkmark  &\checkmark  &\checkmark   &\checkmark \\ \rowcolor{lightblue}
Minimum SE guarantee &$\times$ &$\times$ &$\times$  &$\times$  &$\times$  &$\times$ &\checkmark \\ \rowcolor{lightblue}
\bottomrule
\end{tabular}\label{Contrast_our_work}
}
\end{table}

Besides the DoF analysis, some recent works have studied other performance metrics. 
The authors of \cite{jeonCapacityContinuousspaceElectromagnetic2018} analyzed the capacity of continuous-space electromagnetic (EM) channels, and showed how the capacity is affected by the bandwidth constraint.
In \cite{mikkiShannonInformationCapacity2023}, the authors derived the approximation of the Shannon capacity of continuous surfaces with any given geometries operating in Gaussian channels. 
The authors of \cite{sanguinettiWavenumberDivisionMultiplexingLineofSight2023} proposed a wavenumber-division multiplexing scheme
for a pair of parallel line segments in LoS conditions, which could obtain the same spectral efficiency (SE) as that achieved by the singular value decomposition (SVD) precoding scheme. 
The authors of \cite{wan2023can} studied the mutual information achieved by CAPA and SPDA architectures, and unveiled that the former served as the performance upper bound to the latter without mutual coupling among antennas.
By selecting an optimal basis for both the transmitter and receivers to diagonalize the channel, \cite{iacovelli2025multi} analyzed the SE of  CAPA systems and highlighted that CAPA outperforms SPDA in terms of SE when a finite number of radio-frequency (RF) chains is employed. 

It is worth recalling that beamforming design is essential to leverage the benefits of MIMO and underpin high-quality wireless communications. Apart from the performance analysis, some initial research efforts have focused on the design of CAPA beamformers. For example,
the authors of \cite{zhangPatternDivisionMultiplexingMultiUser2023} developed a Fourier-based approach to maximize the weighted sum rate of CAPA-based multi-user communications, which approximates the continuous CAPA beamformer with a finite number of orthogonal Fourier basis functions. Based on this approximate discretization approach, \cite{qian2024spectral}  studied a multi-user uplink communication system to maximize the SE, where both the base station (BS) and users were equipped with CAPAs. 
Note that the Fourier-based approach can only provide an approximate result and has a potentially high complexity for CAPA beamformer design. To address this issue,  \cite{wang2025beamforming} proposed to employ the calculus of
variations (CoV) theory and directly optimized the CAPA beamformer for maximizing the weighted sum rate of a CAPA-bsaed multi-user communication system. As a further advance, the authors of \cite{wang2025optimalbeamformingmultiusercontinuous} continued to explore the optimal structure of CAPA beamformers and derived closed-form expressions for several typical CAPA beamforming strategies. The obtained results in \cite{wang2025beamforming,wang2025optimalbeamformingmultiusercontinuous} have verified the significant performance gain of CAPAs compared to SPDAs in multi-user communications. Furthermore, the authors of \cite{securebeamformingCAPA2025} explored the applications of CAPAs in secure transmissions, which showed that CAPA-based systems can achieve a higher weighted secrecy sum rate compared to their SDPA counterparts.

\subsection{Motivations and Contributions}
Note that the aforementioned contributions only studied the performance gain of CAPA beamforming in the unicast scenario, where independent data streams are delivered to each user \cite{sanguinettiWavenumberDivisionMultiplexingLineofSight2023,zhangPatternDivisionMultiplexingMultiUser2023,qian2024spectral,wang2025beamforming,wang2025optimalbeamformingmultiusercontinuous,securebeamformingCAPA2025}. 
In such cases, the enhanced spatial DoFs facilitated by a CAPA can be always exploited to mitigate the inter-user interference and thus improve the performance. Nevertheless, there is a dearth of investigations on CAPA-based multicast communications. 
In contrast to unicast, a common data stream is sent to a group of users in multicast scenarios, which makes it especially attractive for applications such as video stream broadcasting or online conferences \cite{9093950}. Moreover, multicast only requires a smaller number of RF chains than unicast, and therefore is more energy- and cost-efficient. Despite these promising benefits, the design of multicast beamforming is different from that of unicast beamforming and is more challenging. In multi-group multicast communications, the beamformer for each group should be designed for not only mitigating the inter-group interference but also maximizing the intra-group worst-case performance. Given the fact that the enhanced spatial resolutions provided by CAPAs will make the users' channel tend to be orthogonal to each other \cite{wang2025beamforming}, an interesting question arises: ``\emph{are CAPAs beneficial for multicast?}''

Motivated by the above question, in this paper, we investigate a CAPA-based multi-group multicast communication system with the aim of maximizing the system energy efficiency (EE). Employing the Dinkelback's method, a CoV-based block coordinate descent (BCD) algorithm and a low-complexity zero-forcing (ZF)-based BCD algorithm are developed for optimizing the CAPA beamformer. Results unveil that \emph{the CAPA architecture still significantly outperforms the SDPA countepart in multi-group multicast communications, while increasing the aperture size is not always beneficial for multicast performance enhancement.}

The main contributions of this paper are summarized below, which are boldly and explicitly compared to the relevant state-of-the-art in \tabref{Contrast_our_work}.
\begin{itemize}
  \item We investigate a CAPA-based multi-group multicast communication system, where the BS employs the continuous source current patterns to transmit a dedicated common information to each user group. We formulate the CAPA beamformer design problem for the maximization of the system EE, subject to each group's minimun multicast SE constraint and the total transmit power constraint. To address this integral-based non-convex optimization problem, we develop a pair of algorithms under the Dinkelbach's framework.  
      
  \item We propose an efficient CoV-based BCD alogorithm. In particular,  we first equivalently transform the non-convex minimum multicast SE constraint into a tractable linear constraint by performing a quadratic transformation. We then employ the BCD algorithm to iteratively solve the reformulated problem,  where the CAPA beamformer design subproblem is optimally solved utilizing the CoV theory and Lagrangian dual method.  Our analysis reveals that the optimized CAPA beamformer is the combinations of all users' channels.
          
  \item We further propose a low-complexity ZF-based approach for CAPA beamformer designs. We select a representative user for each group based on the highest intra-group channel correlation or the lowest inter-group correlation.  Then, we derive the closed-form CAPA beamformer for each multicast group using the selected users' channels with the ZF principle, which greatly reduces the computational complexity. The remaining power allocation subproblem is convex and can be solved efficiently.

  \item Numerical results demonstrate that with the aid of the proposed CoV-based and ZF-based CAPA designs, the CAPA can achieve a higher EE than the conventional SPDA. Different from SPDA, increasing the aperture size of the CAPA might degrade the EE performance and therefore a moderate sized CAPA is sufficient in practice. Moreover, wider user distributions in each multicast group will impose a significant performance degradation on CAPA-based multicast communications.
\end{itemize}

\subsection{Organization and Notations}
The rest of the paper is organized as follows: \secref{Sec:system model and problem formulation} discusses the system model for the CAPA-based multi-group multicast communication system, formulates the EE maximization CAPA beamformer problem, and introduces the Dinkelback's framework. \secref{Sec:CoV-based}  proposes the CoV-based BCD algorithm and \secref{Sec:low complexity} further proposes a low-complexity ZF-based CAPA beamformer design approach. \secref{Sec:simuations} provides simulation results to validate the effectiveness of the proposed algorithms.  Finally,
\secref{Sec:conclusions} concludes the paper.

\emph{Notations:} Scalars, vectors, and matrices are denoted by
lower-case, bold-face lower-case, and bold-face upper-case letters, respectively; 
 $\bX^T$ and $\bX^H$ are the transpose and conjugate transpose of $\bX$, respectively;  $\bx \succeq \bzero$ means each element of $\bx$ is non-negative; $||\bx||$ is the Euclidean norm and $|\bx|$ represents the modulus; calligraphic letters, e.g., $\cG$, denotes sets, with $|\cG|$ being the set cardinality and $\cG\backslash \{g\}$ standing for the exclusion of $g$ from $\cG$;
 $\expect{\cdot}$ and $\trace{\cdot}$ are the expectation operator and trace operator, respectively; $\mathcal{CN}(a,b)$ represents a Gaussian distribution, with $a$ being the mean and $b$ being the variance.

\section{System Model and Problem Formulation}\label{Sec:system model and problem formulation}
In this section, we present the system model for the considered CAPA-based multi-group multicast communication system. Then, we formulate the CAPA beamformer optimization problem to maximize the system EE and introduce the Dinkelbach's framework.

\subsection{System Model}
\begin{figure}[t]
	\includegraphics[width=0.48\textwidth]{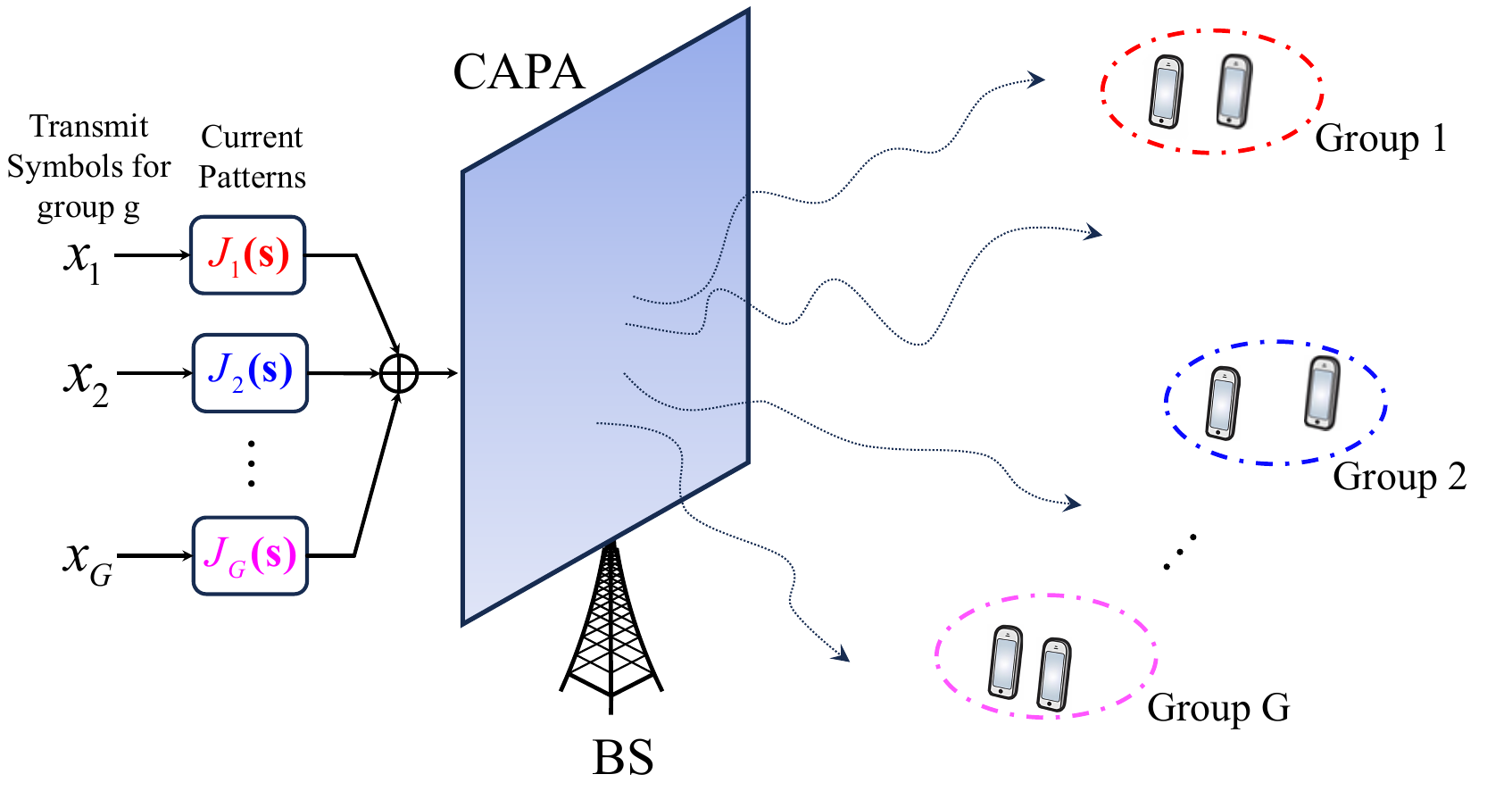}
	\centering
     \captionsetup{font={small}}
	\caption{CAPA-based downlink multi-group multicast communication system.}
    \label{CAPA system}
\end{figure}

As shown in \figref{CAPA system}, we study a downlink CAPA-based multi-group multicast scenario communication system, where a BS is equipped with a CAPA serving $G$ multicast groups. Users in the
same group share the same information data and the information data transmitted for different groups is independent and different,
which means there exists inter-group interference. Without loss of generality, 
the set of groups served by the BS is denoted by $\cG \triangleq \{1,2,\dots,G\}$, with a total number of $|\cG| = G$. 
The number of single-antenna users is denoted by $\cK \triangleq \{1,2,\dots,K\}$, with $|\cK| = K$. It is assumed that the number of users belonging to group $g$ is $|\cK_g| = K_g$, which is denoted by $\cK_g \triangleq \{(g,1),(g,1),\dots,(g,K_g)\} \in \cK$. The sets of users belonging to different groups are disjoint, i.e., $\cK_j \cap \cK_i = \emptyset, \forall i,j \in \cG, i \neq j$, which means that each user is assumed to belong to one group only.

\subsubsection{Transmit Signal}
We assume that the CAPA at the BS, with an area of $S_T = |\cS_{\rT}|$, contains sinusoidal source currents to emit EM waves for wireless communications. The source current density functions for $G$ groups are denoted by $\{\bJ_g(\bs,\omega)\}_{g \in \cG} \in \bbC^{3\times 1}$, with $\bJ_g(\bs,\omega)$ designed for the $g$-th group.
Also, $\bs\in \cS_{\rT} \in \bbR^{3 \times 1}$ is assumed to represent the spatial three-dimensional (3D) coordinates of any point across the aperture of the  CAPA, while $\omega = 2\pi/\lambda$ is the angular frequency. In this paper, we focus on a narrowband single-carrier communication system, where $\omega$ can be omitted. Therefore,  we only denote the source current density function as $\bJ_g(\bs),\forall g\in \cG$.
Accordingly, $\{\bJ_g(\bs)\}_{\forall g\in \cG}$ can be expressed as follows:
\begin{align}
  \bJ_g(\bs) = J_{g,x}(\bs)\hat{\bu}_x + J_{g,y}(\bs)\hat{\bu}_y + J_{g,z}(\bs)\hat{\bu}_z,
\end{align}
where $\hat{\bu}_i\in \bbR^{3 \times 1},\forall i\in \{x,y,z\}$ represents the unit vector along the $i$-axis.
In this paper, we focus on the case of a vertically-polarized transmitter and only the $y$-component of the source current is excited. Thus, we have
 $ \bJ_g(\bs) = J_g(\bs)\hat{\bu}_y $.
Here, we use $J_g(\bs) \triangleq J_{g,y}(\bs)$ to simplify the source current expression.

For the considered multi-group multicast transmission, the transmit signal at the BS is given by
\begin{align}
  J(\bs) = \sum_{g = 1}^{G} J_g(\bs)x_g,
\end{align}
where $J_g(\bs) \in \bbC$ is the source current pattern and $x_g \in \bbC$ is the transmit symbol for the $g$-th group, which satisfies $\expect{\bx\bx^H} = \bI_G$, with $\bx = [x_1,x_2,\dots,x_G]^T$.

\subsubsection{Receive Signal}
Let $\br_{g,k}\in \bbR^{3\times 1},\forall g \in \cG, \forall k \in \cK_g$ denote the spatial 3D coordinates of user $k$ in group $g$. According to Maxwell's equations, the electric field generated by the source current function $\bJ_g(\bs)$ at the position of $\br_{g,k}$ in a homogeneous medium is given by \cite{sanguinettiWavenumberDivisionMultiplexingLineofSight2023}
\begin{align}
  \be_{g,k}(\br) = \int_{\cS_{\rT}} \bG(\br_{g,k},\bs)\bJ_g(\bs) d\bs \in \bbC^{3\times1},
\end{align}
where $\bG(\br,\bs)$ is the Green's function for the LoS scenario, which can be expressed as follows:
\begin{align}\label{Green}
  \bG(\br,\bs) = -\frac{j \eta e^{-j \frac{2\pi}{\lambda} ||\br-\bs||}}{2\lambda||\br-\bs||}\left(\bI_3 - \frac{(\br-\bs)(\br-\bs)^T}{||\br-\bs||^2}\right).
\end{align}
Here, $\eta$ is the free-space intrinsic impedance. We assume that each user is equipped with a uni-polarized antenna, where the polarization direction is denoted by $\hat{\bu}_{g,k} \in \bbC^{3 \times 1}$ and $||\hat{\bu}_{g,k}|| = 1$. As a result, the user only captures the component of the electric field $\be_{g,k}(\br)$ along the direction of $\hat{\bu}_{g,k}$. Thus, user $(g,k)$ receives the following noisy electric field:
\begin{align}
  y_{g,k} =& \int_{\cS_{\rT}} h_{g,k}(\bs)J_g(\bs) x_g d\bs \notag \\
  &+ \overbrace{\sum_{i \in \cG \backslash \{g\}}\int_{\cS_{\rT}} h_{g,k}(\bs)J_i(\bs)x_i d\bs}^{\text{inter-group interference}} + n_{g,k},
\end{align}
where $h_{g,k}(\bs) = \hat{\bu}_{g,k}^T \bG(\br_{g,k}, \bs)\hat{\bu}_y \in \bbC$ represents the continuous EM channel from the BS to user $(g,k)$, while
$n_{g,k} \sim \CN(0,\sigma_{g,k}^2)$ is the additive white Gaussian noise. 

\subsubsection{System EE}
For user $(g,k)$, $\forall k \in \cK_g, \forall g \in \cG$, the received signal-to-interference-plus-noise-ratio (SINR) is given by
\begin{align}
  \gamma_{g,k} =& \frac{\left|\int_{\mathcal{S}_{\mathrm{T}}} h_{g,k}(\bs) J_g(\bs) d \mathbf{s}\right|^2}{\sum_{i \in \cG \backslash \{g\}} \left|\int_{\mathcal{S}_{\mathrm{T}}} h_{g,k}(\bs) J_i(\bs) d \mathbf{s}\right|^2+\sigma_{g,k}^2}.
\end{align}
The achievable SE (in bit/s/Hz) is given by
\begin{align}
  R_{g,k}(\bJ)= \log_2 \left( 1+\gamma_{g,k}(\bJ) \right),\forall k \in \cK_g, \forall g \in \cG,
\end{align}
where $\bJ \triangleq \{J_g(\bs)\}_{g=1}^{G}$.
Due to the nature of the multicast mechanism, the achievable SE of group $g$ is limited by the minimum user SE in this
group, which is defined as follows:
\begin{align}\label{R_J}
  R_g(\bJ) \triangleq \min_{k \in \cK_g}\log_2 \left( 1+\gamma_{g,k}(\bJ) \right), \forall g\in \cG.
\end{align}
The total transmit power consumption is modeled as
\begin{align}\label{power_orig}
  P(\bJ) = \sum_{g=1}^G \int_{\mathcal{S}_{\mathrm{T}}}\left|J_g(\bs)\right|^2 d \mathbf{s}.
\end{align}
As a result, the EE of this multi-group multicast system is given by
\begin{align}
{\rm EE} = \frac{R(\bJ)}{P(\bJ)} \triangleq \frac{\sum_{g=1}^{G} R_g(\bJ)}{\sum_{g=1}^G \int_{\mathcal{S}_{\mathrm{T}}}\left|J_g(\bs)\right|^2 d \mathbf{s}}.
\end{align}

\subsection{Problem Formulation}
In this paper, we aim to design the CAPA beamformer to maximize the EE for the considered multi-group multicast system. The corresponding optimization problem can be formulated as follow:
\begin{subequations}\label{EE_problem_ori}
  \begin{align}
      \max _{\bJ} \quad &  {\rm EE}\\
     \st \quad & \min_{k \in \cK_g}\log_2\left( 1+\gamma_{g,k}(\bJ)\right) \geq  \bar{R}_{g},\quad \forall g \in \cG \label{QoS_guarantee},\\
         & \sum_{g=1}^G \int_{\mathcal{S}_{\mathrm{T}}}\left|J_g(\bs)\right|^2 d \mathbf{s} \leq P_t.\label{power_constraint_orig}
  \end{align}
\end{subequations}
 Constraint \eqref{QoS_guarantee} is imposed to guarantee that the date SE of user $(g,k),\forall g \in \cG,\forall k \in \cK_g$ achieves the minimum multicast SE requirement of group $g$, which is denoted by $\bar{R}_{g}$, while constraint \eqref{power_constraint_orig} ensures that the transmit power is within the power budget, $P_t$.

\subsection{Dinkelbach's Framework}
Considering that the EE maximization problem \eqref{EE_problem_ori} is a non-convex fractional programming problem, we employ the Dinkelbach's method to transform it into a non-fractional form. Denote the optimized solution of \eqref{EE_problem_ori} and the resulting maximized EE as $\bJ^{\rm opt}$ and $\eta^{\rm opt} = \frac{R(\bJ^{\rm opt})}{P(\bJ^{\rm opt})}\triangleq \underset{\bJ}{\max} \frac{ R(\bJ)}{P(\bJ)}$, respectively. It can be observed that finding the optimized solution of \eqref{EE_problem_ori} is equivalent to finding the optimized $\eta^{\rm opt}$. A two-layer algorithm framework is adopted to find the optimized $\eta^{\rm opt}$. Initialize $\eta$ with a small value. In each iteration, $\eta$ remains constant in the inner layer and is updated in the outer layer until convergence is reached.
 
The subproblem at the $\ell$-th iteration of Dinkelbach's algorithm can be reformulated as follows,
\begin{subequations}\label{Dinkel_problem}
  \begin{align}
 \max_{\bJ} \quad & R(\bJ) - \eta^{(\ell)} P(\bJ) \label{Dinkel_obj}\\
\st\quad & \eqref{QoS_guarantee}, \eqref{power_constraint_orig},
  \end{align}
\end{subequations}
where $R(\bJ)$ and $P(\bJ)$ were given in \eqref{R_J} and \eqref{power_orig}, respectively. The auxiliary variable is updated by
\begin{align}\label{Dinkel_update}
  \eta^{(\ell)} = \frac{R(\bJ^{(\ell)})}{P(\bJ^{(\ell)})}.
\end{align}

Despite the non-fractional form of problem \eqref{Dinkel_problem}, it is still difficult to be globally optimally solved due to two main reasons. 
First, the continuous characteristics of the optimizing variables $\bJ$ and the integral-based problem may result in high computational complexity.
Second, the non-convexity of the objective function and the minimum multicast SE constraint imposed on each group make the problem non-trivial to handle.  In the following, we will develop a pair of BCD algorithms to solve problem \eqref{Dinkel_problem}.

\section{CoV-based BCD Algorithm}\label{Sec:CoV-based}
In this section, we propose an efficient CoV-based BCD algorithm to solve the EE maximization problem \eqref{Dinkel_problem}  within the framework of Dinkelbach's method.

\subsection{Problem Reformulation}
Letting $r_g \triangleq \min_{k \in \cK_g} \gamma_{g,k}(\bJ),\forall g \in \cG$,
we can rewrite problem \eqref{Dinkel_problem} equivalently as follows:
\begin{subequations}\label{CP_with_r_gk}
  \begin{align}
    \max _{\bJ,\br}\quad & R(\br) - \eta^{(\ell)}P(\bJ)\\
     \st \quad & r_{g} \leq \frac{\left|\int_{\mathcal{S}_{\mathrm{T}}} h_{g,k}(\bs) J_g(\bs) d \mathbf{s}\right|^2}{\sum_{i \in \cG \backslash \{g\}} \left|\int_{\mathcal{S}_{\mathrm{T}}} h_{g,k}(\bs) J_i(\bs) d \mathbf{s}\right|^2+\sigma_{g,k}^2},\notag\\
     & \forall g \in \cG, \forall k \in \cK_g, \label{SINR_constraint}\\
         & r_{g} \geq  2^{\bar{R}_g}-1,\forall g \in \cG,\label{rate_constraint}\\
         & \sum_{g=1}^G \int_{\mathcal{S}_{\mathrm{T}}}\left|J_g(\bs)\right|^2 d \mathbf{s} \leq P_t\label{power_constraint},
  \end{align}
\end{subequations}
where $\br \triangleq \{r_{g}\}_{g\in \cG}$ and $R(\br)= \sum_{g \in \cG} \log_2\left(1+r_g\right)$.

Note that in problem \eqref{CP_with_r_gk}, the presence of non-convex constraint \eqref{SINR_constraint} makes it difficult to address. To cope with the problem, we first introduce the following lemma.
\begin{lemma}\label{mu_k_opt}
For  given  $\{h_{g,k}(\bs)\}_{\forall g \in \cG, \forall k \in \cK_g}$, the problem \eqref{CP_with_r_gk} is equivalent to the following one:
\begin{subequations}\label{Before_BCD_problem}
\begin{align}
  \max _{\bJ,\br,\bmu}\quad & R(\br) - \eta^{(\ell)}P(\bJ)\label{Din_ori_obj}\\
  \st\quad & r_g \leq y(\mu_{g,k},\bJ), \forall g \in \cG, \forall k \in \cK_g,\label{equivalent_r_g_constraint}\\
  & \eqref{rate_constraint},\eqref{power_constraint}\notag,
\end{align}
\end{subequations}
where $y(\mu_{g,k},\bJ)$ is given by \eqref{y_def} at the top of the next page, while $\bmu = \{\mu_{g,k}\}_{\forall g \in \cG, \forall k \in \cK_g}$ are auxiliary variables. 
\begin{figure*}
\begin{align}\label{y_def}
 y(\mu_{g,k},\bJ) = 2\real{\mu_{g,k}^* \int_{\mathcal{S}_{\mathrm{T}}} h_{g,k}(\bs) J_g(\bs) d \mathbf{s}}  -|\mu_{g,k}|^2 \left(\sum_{i \in \cG \backslash \{g\}} \left|\int_{\mathcal{S}_{\mathrm{T}}} h_{g,k}(\bs) J_i(\bs) d \mathbf{s}\right|^2+\sigma_{g,k}^2\right).
\end{align}
\end{figure*}
For each $\mu_{g,k}$, its optimal value is given by
\begin{align}
\mu_{g,k}^{\rm opt} = \frac{\int_{\mathcal{S}_{\mathrm{T}}} h_{g,k}(\bs) J_g(\bs) d \mathbf{s}}{\sum_{i \in \cG \backslash \{g\}} \left|\int_{\mathcal{S}_{\mathrm{T}}} h_{g,k}(\bs) J_i(\bs) d \mathbf{s}\right|^2+\sigma_{g,k}^2}. \label{opt_eta_k}
\end{align}
\end{lemma}

\begin{proof}
For any given $\bJ$, the right-hand expression $y(\mu_{g,k},\bJ)$ can be reduced to $y(\mu_{g,k})$. 
Since $r_g$ represents the minimum multicast SE of group $g$, it should be as large as possible at the optimized solutions of problem \eqref{Before_BCD_problem}. Note  each $r_g$ is upper bounded by $K_g$ terms $\{y(\mu_{g,k})\}_{\forall k \in \cK_g}$, which are independent of each other. Therefore, we can equivalently write constraint \eqref{equivalent_r_g_constraint} as follows:
\begin{align}
  r_g \leq \max_{\mu_{g,k}}  \: y(\mu_{g,k}) \triangleq y_{\max}(\mu_{g,k}),\quad \forall k \in \cK_g,\forall g \in \cG.
\end{align}
It can be observed that each $y(\mu_{g,k})$ is a concave function over $\mu_{g,k}$ and it attains its maximal value when its first-order derivative is zero,  at which point the optimal $\bmu^{\rm opt} \triangleq \{\mu^{\rm opt}_{g,k}\}_{\forall g \in \cG,\forall k\in \cK_g}$ is given by \eqref{opt_eta_k}.
Substituting \eqref{opt_eta_k} into $y(\mu_{g,k})$, we have the maximal value of $y_{\max}(\mu_{g,k})$ as follows:
\begin{align}
  y_{\max}(\mu_{g,k}^{\rm opt}) =& \frac{\left|\int_{\mathcal{S}_{\mathrm{T}}} h_{g,k}(\bs) J_g(\bs) d \mathbf{s}\right|^2}{\sum_{i \in \cG \backslash \{g\}} \left|\int_{\mathcal{S}_{\mathrm{T}}} h_{g,k}(\bs) J_i(\bs) d \mathbf{s}\right|^2+\sigma_{g,k}^2},
\end{align}
 which recovers the right-hand side of \eqref{SINR_constraint}. This completes the proof.
\end{proof}

With the reformulation, problem \eqref{Before_BCD_problem} has a more tractable form with the optimization variables $\{\bJ,\br,\bmu\}$, which are coupled. In the following, we will employ the BCD method to alternatively optimize the variables.

\subsection{BCD Algorithm}
For the joint optimization problem \eqref{Before_BCD_problem}, we divide the optimization variables into two blocks: $\{\bJ,\br\}$ and $\{\br,\bmu\}$, and employ the BCD algorithm to iteratively solve each subproblem as follows.
\subsubsection{Subproblem with respect to $\bJ$ and $\br$}For given $\bmu$, the CAPA beamformer design subproblem with respect to $\bJ$ and $\br$ is given by
  \begin{align}\label{subproblem_J_r}
   \max_{\bJ,\br} \quad &  \sum_{g=1}^{G} \log_2\left(1+r_g\right)- \eta^{(\ell)}\sum_{g=1}^G \int_{\mathcal{S}_{\mathrm{T}}}\left|J_g(\bs)\right|^2 d \mathbf{s} \\
\st\quad & \eqref{rate_constraint},\eqref{power_constraint},\eqref{equivalent_r_g_constraint}\notag.
  \end{align}
As the problem \eqref{subproblem_J_r} is jointly convex with respect to $\bJ$ and $\br$, we employ the Lagrangian dual method and the CoV theory to optimally solve it. 

The partial Lagrangian function of problem \eqref{subproblem_J_r} is given in \eqref{lagrangian_dual_problem} at the top of the next page,
\begin{figure*}
\begin{subequations}\label{lagrangian_dual_problem}
\begin{align}
  \cL\left(\bJ, \br, \blambda,\xi\right)=& \sum_{g=1}^{G}\sum_{k=1}^{K_g} \lambda_{g,k} \left[2\real{\mu_{g,k}^* \int_{\mathcal{S}_{\mathrm{T}}} h_{g,k}(\bs) J_g(\bs) d \mathbf{s}} -|\mu_{g,k}|^2 \left(\sum_{i \in \cG \backslash \{g\}} \left|\int_{\mathcal{S}_{\mathrm{T}}} h_{g,k}(\bs) J_i(\bs) d \mathbf{s}\right|^2 + \sigma_{g,k}^2\right)- r_g \right]\notag \\
  &+ \xi\left(P_t - \sum_{g=1}^G \int_{\mathcal{S}_{\mathrm{T}}}\left|J_g(\bs)\right|^2 d \mathbf{s} \right)+\sum_{g=1}^{G} \log_2\left(1+r_g\right)-\eta^{(\ell)}\sum_{g=1}^G \int_{\mathcal{S}_{\mathrm{T}}}\left|J_g(\bs)\right|^2 d \mathbf{s} \\
  & \triangleq \sum_{g=1}^{G}\left[f_1(r_g) + f_2\left(J_g(\bs)\right)\right] +\Sigma_1. 
\end{align}
\end{subequations}
\hrulefill
\end{figure*}
where
\begin{subequations}
\begin{align}
& f_1(r_g) =  \log_2\left(1+r_g\right) -  r_g \sum_{k=1}^{K_g} \lambda_{g,k}, \\
& f_2(J_g(\bs)) =  \sum_{k=1}^{K_g}2 \lambda_{g,k}  \real{\mu_{g,k}^* \int_{\mathcal{S}_{\mathrm{T}}} h_{g,k}(\bs) J_g(\bs) d \mathbf{s}} \notag \\
&- \sum_{g' \neq g}^{G}\sum_{k=1}^{K_{g'}}\lambda_{g',k} |\mu_{g',k}|^2\left|\int_{\mathcal{S}_{\mathrm{T}}} h_{g',k}(\bs) J_{g}(\bs) d \mathbf{s}\right|^2 \notag \\
&- (\xi+\eta^{(\ell)}) \int_{\mathcal{S}_{\mathrm{T}}}\left|J_g(\bs)\right|^2 d \mathbf{s}\label{f_2},\quad \forall g \in \cG,\\
& \Sigma_1 = \xi P_t - \sum_{g=1}^{G} \sum_{k=1}^{K_g} \lambda_{g,k} |\mu_{g,k}|^2 \sigma_{g,k}^2.
\end{align}
\end{subequations}
Note that $\xi $ and $\blambda\triangleq \{\lambda_{g,k}\}_{\forall g \in \cG, \forall k \in \cK_g}$ are the non-negative Lagrangian multipliers associated with constraints  \eqref{power_constraint} and \eqref{equivalent_r_g_constraint}, respectively.

Then, the partial Lagrange dual function is 
\begin{align}\label{Lag_dual_problem}
  g(\blambda,\xi) = \max_{\bJ,r_g \geq 2^{\bar{R}_g}-1,\forall g \in \cG} \: \cL\left(\bJ, \br, \blambda,\xi\right).
\end{align}
Accordingly, the dual problem of \eqref{subproblem_J_r} can be expressed as
\begin{align}\label{dual_problem}
  \max_{\blambda \succeq 0,\xi\geq 0}  g(\blambda,\xi).
\end{align}
Since  the original problem \eqref{subproblem_J_r} is convex,  we can obtain the optimal solutions of it
by solving the Lagrange dual problem \eqref{dual_problem}. The detailed solving process is clarified as follows:
\paragraph{Obtaining $g(\blambda,\xi)$ via Solving Problem \eqref{Lag_dual_problem}}For given dual variables $\blambda$ and $\xi$,
$g(\blambda,\xi)$ can be obtained by solving the problem \eqref{Lag_dual_problem}, the solutions of which are, respectively, given in \propref{optimal_rg_prop} and \propref{optimal_Jg_prop}.  

\begin{proposition}\label{optimal_rg_prop}
For given $\blambda$ and $\xi$, the optimal $\br^{\rm opt}\triangleq \{r_g^{\rm opt}\}_{\forall g \in \cG}$ to problem \eqref{Lag_dual_problem} is given by
  \begin{align}\label{opt_r_g}
  &r_g^{\rm opt}
  \left\{\begin{aligned}
  &2^{\bar{R}_g} - 1, \quad  2^{\bar{R}_g} - 1 \geq  \frac{1}{\ln 2 \sum_{k=1}^{K_g}\lambda_{g,k}} - 1,&\\
   &\frac{1}{\ln 2 \sum_{k=1}^{K_g}\lambda_{g,k}} - 1, \quad \quad \quad\quad  \quad\text{otherwise.}&
  \end{aligned} \right.
 \end{align}
\end{proposition}

\begin{proof}
The only term of $\cL\left(\bJ, \br, \blambda,\xi\right)$ that contains $\br^{\rm opt}$  is $\{f_1(r_g)\}_{\forall g \in \cG}$, which is continuous and concave with respect to $\br$. Note that the range of $r_g$ is $[2^{\bar{R}_g}-1,\infty)$, thus, at the maximum of $f_1(r_g)$, $r_g$ proves to be either at the stationary point, or the boundary point.
At the stationary point, the first-order derivative of $f_1(r_g)$ with respect to $r_g$ must be 0 for given $\blambda$ and $\xi$, which leads to
\begin{align}\label{optimal_rg}
   \frac{\partial f_1(r_g)}{\partial r_g} = 0 \Rightarrow r_g^{\rm opt} = \frac{1}{\ln 2 \sum_{k=1}^{K_g}\lambda_{g,k}} - 1,\forall g\in \cG,
\end{align}
while at the boundary point, $r_g^{\rm opt} = 2^{\bar{R}_g}-1$. This completes the proof.
\end{proof}
 
 Before presenting \propref{optimal_Jg_prop}, we introduce the following theorem first.
 \begin{theorem}\label{optimal_structure_Jg_theorem}
  The function $f_2(J_g(\bs))$ in \eqref{f_2} achieves its maximum when $\bJ$ takes the following form:
  \begin{align}\label{optimal_tilde_J}
   & J_g(\bs) = \frac{\sum_{k=1}^{K_g} \lambda_{g,k} \mu_{g,k} h_{g,k}^*(\bs)}{\xi+\eta^{(\ell)}} - \notag\\
   & \frac{\sum_{g' \neq g}^{G}\sum_{k=1}^{K_{g'}}\lambda_{g',k}|\mu_{g',k}|^2 h_{g',k}^*(\bs)\int_{\mathcal{S}_{\mathrm{T}}} h_{g',k}(\bz)\tilde{J}_g(\bz) d \bz}{\xi+\eta^{(\ell)}}.
  \end{align}
\end{theorem}
\begin{proof}
At any local maximum point $\tilde{\bJ}\triangleq \{\tilde{J}_g(\bs)\}_{\forall g \in \cG}$, of $\{f_2(J_g(\bs))\}_{\forall g\in \cG}$, for any $\delta \rightarrow 0$, the following inequality holds:
\begin{align}
  f_2\left(\tilde{J}_g(\bs)\right) \geq f_2\left(\tilde{J}_g(\bs)+\delta U_g(\bs)\right),\quad \forall g\in \cG,
\end{align}
where $U_g(s)$ is any smooth function satisfying $U_g(\bs) = 0, \forall \bs \in \partial \mathcal{S}_{\mathrm{T}}$ \cite{wang2025beamforming}, while $\delta U_g(\bs)$ is the variation of $\tilde{J}_g(\bs)$.
Let $\Psi(\delta)\triangleq f_2\left(\tilde{J}_g(\bs)+\delta U_g(\bs)\right)$, which is shown in \eqref{CoV_Ug} at the top of the next page,
\begin{figure*}[htbp]
\begin{align}\label{CoV_Ug}
&\Psi(\delta) =   \sum_{k=1}^{K_g} 2\delta\lambda_{g,k} \real{\mu_{g,k} \int_{\mathcal{S}_{\mathrm{T}}} U_g^*(\bs) h_{g,k}^*(\bs) d \mathbf{s}}
 -(\xi+\eta^{(\ell)}) \int_{\mathcal{S}_{\mathrm{T}}} \left(\delta^2|U_g(\bs)|^2 + 2\delta\real{\tilde{J}_g(\bs)U_g^*(\bs) } \right)d \mathbf{s}+ \Sigma_2\notag \\
 &-\sum_{g' \neq g}^{G}\sum_{k=1}^{K_{g'}}\lambda_{g',k}|\mu_{g',k}|^2 \real{\int_{\mathcal{S}_{\mathrm{T}}} \int_{\mathcal{S}_{\mathrm{T}}}\left[2\delta U_g^*(\bs_1)h_{g',k}^*(\bs_1)h_{g',k}(\bs_2)
 \tilde{J}_g(\bs_2) + \delta^2 U_g^*(\bs_1)  U_g(\bs_2) h_{g',k}^*(\bs_1)h_{g',k}(\bs_2)\right] d\bs_1 d \bs_2}.
\end{align}
\end{figure*}
where $\Sigma_2$ is a constant independent of $\delta$.
Since $f_2(J_g(\bs))$ achieves its local maximum at $\tilde{J}_g(\bs)$, it follows that $\Psi(\delta)$ attains its maximum at $\frac{d \Psi(\delta)}{d \delta}\big|_{\delta = 0} = 0$, leading to the following condition:
  \begin{align}
    &\real{\int_{\mathcal{S}_{\mathrm{T}}}U_g^*(\bs) V_g(\bs)} = 0, \label{J_g_opt_main}
  \end{align}
where $V_g(\bs)$ is given in \eqref{Vg} at the top of the next page.
\begin{figure*}
  \begin{align}\label{Vg}
  V_g(\bs) =   \sum_{k=1}^{K_g} 2\lambda_{g,k} \mu_{g,k} h_{g,k}^*(\bs) - 2(\xi+\eta^{(\ell)}) \tilde{J}_g(\bs)- \sum_{g' \neq g}^{G}\sum_{k=1}^{K_{g'}}2\lambda_{g',k}|\mu_{g',k}|^2 h_{g',k}^*(\bs)\int_{\mathcal{S}_{\mathrm{T}}} h_{g',k}(\bz)\tilde{J}_g(\bz) d \bz.
  \end{align}
  \hrulefill
\end{figure*}
According to \cite[Lemma 3]{wang2025beamforming}, $V_g(\bs) = 0$.
Thus,  we obtain the optimal structure of $J_g(\bs)$ shown in \eqref{optimal_tilde_J}, which completes the proof.
\end{proof}
Then, we have the following proposition to derive the optimal $\bJ^{\rm opt}\triangleq \{J_g^{\rm opt}(\bs)\}_{\forall g \in \cG}$.
 \begin{proposition}\label{optimal_Jg_prop}
Since $\cL\left(\bJ, \br, \blambda,\xi\right)$ is concave with respect to $\bJ$, at the maximum of  $\cL\left(\bJ, \br, \blambda,\xi\right)$, the optimal CAPA beamformer $\bJ^{\rm opt}$, is
   \begin{align}\label{optimal_Jg}
   J^{\rm opt}_g(\bs)
    =  \sum_{l \in \cK_g} p_l g_l^*(\bs) - \sum_{l \in \cK_g} \sum_{j,i\in \cK \backslash\{\cK_g\}}  p_l p_i c_{j,i} g_j^*(\bs) \langle g_i,g_l\rangle,
  \end{align}
  where  $p_i =\frac{ \lambda_i}{\xi+\eta^{(\ell)}}$ and $ g_i(\bs) =  \mu_i^* h_i(\bs)$.
 \end{proposition}

\begin{proof}
By utilizing \theoref{optimal_structure_Jg_theorem}, we can obtain the optimal structure of $\bJ$ when the first-order partial derivative of $\cL\left(\bJ, \br,\blambda,\xi\right)$ with respect to  $\bJ$ equals $0$. 
It is worth noting that  we have only obtained an integral equation, i.e., \eqref{optimal_tilde_J}, that the optimal $\bJ$ should satisfy.

To extract $\tilde{\bJ}$ from the integral equation \eqref{optimal_tilde_J},  we first rewrite the expression of \eqref{optimal_tilde_J} as follows:
\begin{align}
\int_{\mathcal{S}_{\mathrm{T}}} I(\bs,\bz)  \tilde{J}_g(\bz) d\bz   = \sum_{k=1}^{K_g}\frac{ \lambda_{g,k} \mu_{g,k} h_{g,k}^*(\bs)}{\xi+\eta^{(\ell)}},\:\forall g \in \cG,
\end{align}
where
\begin{align}\label{I}
  & I(\bs,\bz) = \delta(\bs-\bz)  \notag \\
  &+ \frac{1}{\xi+\eta^{(\ell)}}\sum_{g' \neq g}^{G}\sum_{k=1}^{K_{g'}}\lambda_{g',k}|\mu_{g',k}|^2 h_{g',k}^*(\bs) h_{g',k}(\bz).
\end{align}

If we can find the inversion of $I(\bs,\bz)$, which can be denoted as $I^{-1}(\bs',\bs)$, then we can obtain $\tilde{J}_g(\bs')$ as
\begin{subequations}\label{J_g_derivation}
  \begin{align}
 & \int_{\mathcal{S}_{\mathrm{T}}} I^{-1}(\bs',\bs) \int_{\mathcal{S}_{\mathrm{T}}} I(\bs,\bz)  \tilde{J}_g(\bz) d\bz d \bs \\
&= \int_{\mathcal{S}_{\mathrm{T}}} \delta(\bs'-\bz) \tilde{J}_g(\bz) d\bz \\
  &= \tilde{J}_g(\bs'),
\end{align}
\end{subequations}
where $\delta(\cdot)$ is the Dirac delta function.

Let $i \triangleq [(g-1) K_g+k]$.
Then, \eqref{I} can be rewritten as follows:
\begin{subequations}\label{J_g_optimal}
\begin{align}
  I(\bs,\bz) =& \delta(\bs-\bz) + \frac{1}{\xi+\eta^{(\ell)}}\!\!\!
    \sum_{i \in \cK\backslash\{\cK_g\}}\!\! \lambda_{i}|\mu_{i}|^2 h_{i}^*(\bs) h_{i}(\bz),\\
    =& \delta(\bs-\bz) + \sum_{i \in \cK\backslash\{\cK_g\}}p_i g_{i}^*(\bs) g_{i}(\bz).
\end{align}
\end{subequations}

According to \cite[Lemma 2]{wang2025optimalbeamformingmultiusercontinuous}, the structure of the inverse of $I(\bs,\bz)$ can be expressed as
\begin{align}\label{inverse_Jg_structure}
  I^{-1}(\bs',\bs) = \delta(\bs'-\bs) - \sum_{j,i\in \cK\backslash\{\cK_g\}} p_i  c_{j,i} g_j^*(\bs')g_i(\bs),
\end{align}
so that
\begin{align}\label{inver_Jg_times_Jg}
  \int_{\cS_T} I^{-1}(\bs',\bs) I(\bs,\bz) d \bs = \delta(\bs'-\bz).
\end{align}
Thus, the remaining problem is to find the coefficients $\{c_{j,i}\}$, the process of which is clarified in the following.

By substituting \eqref{J_g_optimal} and \eqref{inverse_Jg_structure} into \eqref{inver_Jg_times_Jg}, we have the equation in \eqref{inv_I_equation} at the top of the next page,
\begin{figure*}
\begin{align}\label{inv_I_equation}
 \int_{\cS_T} I^{-1}(\bs',\bs) I(\bs,\bz) d \bs  = \delta(\bs'-\bz) + \!\!\!\!\sum_{v\in \cK \backslash \{\cK_g\}}p_v g_v^*(\bs') g_v(\bz) - \!\!\!\!\sum_{j,i\in \cK \backslash \{\cK_g\}}\!\!\!\! p_i c_{j,i} g_j^*(\bs')g_i(\bz) -\!\!\!\!\!\!\!\!\sum_{j,i,v\in \cK \backslash \{\cK_g\}}p_i p_v c_{j,i} g_j^*(\bs') g_v(\bz)\langle g_i,g_v \rangle.
\end{align}
\hrulefill
\end{figure*}
where
\begin{align}\label{channel_correlations}
  \langle g_i,g_v \rangle = \int_{\mathcal{S}_{\mathrm{T}}} g_i(\bs) g_v^*(\bs) d \bs.
\end{align}
Since the inverse of $I(\bs,\bz)$, i.e., $I^{-1}(\bs',\bs)$, must satisfy the relationship shown in \eqref{inv_I_equation},
 the last three sum terms of \eqref{inv_I_equation} should equal 0. 
Thus, we can obtain the coefficient of $c_{j,i}$ as follows:
\begin{align}
c_{j,i} = [(\bI_{K-K_g}+\bP_{\rm oth} \bQ^T)^{-1}]_{j,i}\label{cal_c_ji},
\end{align}
where $\bP_{\rm oth} = \rdiag{p_1,\dots,p_i,\dots},\forall i \in \cK\backslash \{\cK_g\}$, while the $(i,j)$-th element of $\bQ$ is
\begin{align}\label{cal_Q}
  [\bQ]_{i,j} = \langle g_i,g_j \rangle, \forall i,j \in \cK\backslash \{\cK_g\}.
\end{align}
Substituting $I^{-1}(\bs',\bs)$ into \eqref{J_g_derivation}, we can obtain the optimal $J_g(\bs),\forall g \in \cG$ shown in \eqref{optimal_Jg}, which completes the proof.
\end{proof}

The first term in the expression of \eqref{optimal_Jg} represents the signal strength of users within the group, while the second term of \eqref{optimal_Jg} is the inter-group interference. This observation reveals that the optimized CAPA beamformer in multicast scenarios  also lies within the subspace spanned by the channel responses of all users in the current system, which aligns with the results in \cite{wang2025optimalbeamformingmultiusercontinuous} with unicast scenarios.

\paragraph{Finding Optimal Dual Solutions $\blambda^{\rm opt}$ and $\xi^{\rm opt}$}
With the obtained optimal solutions $\{\bJ^{\rm opt},\br^{\rm opt}\}$ for given $\{\blambda,\xi\}$, we employ sub-gradient based methods, e.g., ellipsoid method \cite{ellipsoid}, to iteratively solve the dual problem \eqref{dual_problem}. In each iteration, the used sub-gradient for updating $\{\blambda,\xi\}$ is denoted by $(\Delta \blambda,\Delta \xi)$, which is given by
\begin{subequations}\label{derivative_Lag}
  \begin{align}
  & \Delta \lambda_{g,k} = 2\real{{\mu}_{g,k}^* \int_{\mathcal{S}_{\mathrm{T}}} h_{g,k}(\bs) J_g^{\rm opt}(\bs) d \mathbf{s}}  -r_g^{\rm opt} \notag \\
  &-|{\mu}_{g,k}|^2 \left(\sum_{i \in \cG \backslash \{g\}} \left|\int_{\mathcal{S}_{\mathrm{T}}} h_{g,k}(\bs) J_i^{\rm opt}(\bs) d \mathbf{s}\right|^2 + \sigma_{g,k}^2\right),\label{grad_lambda}\\
   &  \Delta \xi = P_t - \sum_{g=1}^G \int_{\mathcal{S}_{\mathrm{T}}}\left|J_g^{\rm opt}(\bs)\right|^2 d \mathbf{s}\label{grad_xi}.
   \end{align}
\end{subequations}
Using the above sub-gradients, the dual variables can be updated by the ellipsoid method towards the optimal solutions, which are denoted by $\blambda^{\rm opt}$ and $\xi^{\rm opt}$.

\paragraph{Obtaining Optimal Primal Solutions to Problem \eqref{subproblem_J_r}} By substituting the optimal dual solutions $\{\blambda^{\rm opt},\xi^{\rm opt}\}$ into \eqref{opt_r_g} and \eqref{optimal_Jg}, we can obtain the optimal primal solutions $\{\bJ^{\rm opt},\br^{\rm opt}\}$.

The algorithm to solve the subproblem  for the optimization of $\bJ$ and $\br$ is summarized in \alref{alg:ellipsoid}.
\begin{algorithm}[htbp]
\caption{Algorithm for Solving Subproblem \eqref{subproblem_J_r}}
\label{alg:ellipsoid}
\begin{algorithmic}[1]
\Require Initial ellipsoid $\cE \left(\bQ, \bx \right)$ containing $(\blambda^{\rm opt},\xi^{\rm opt})$, where $\bx \triangleq [\lambda_{1,1},\dots,\lambda_{g,k},\dots,\lambda_{G,K_g},\xi]^T$ is the center point of $\cE \left(\bQ, \bx \right)$ and the positive definite matrix $\bQ$ characterizes the size of $\cE \left(\bQ, \bx \right)$.
\Repeat
\State Obtain $\{\br^{\rm opt},\bJ^{\rm opt}\}$ using \propref{optimal_rg_prop} and \propref{optimal_Jg_prop};
\State Compute the subgradients of $g(\blambda,\xi)$ using \eqref{grad_lambda} and \eqref{grad_xi}, and update $(\blambda,\xi)$ via ellipsoid method;
\Until{$(\blambda,\xi)$ converges with a prescribed accuracy.} 
\Ensure Calculate the optimal $\{\bJ^{\rm opt},\br^{\rm opt}\}$ by substituting $(\blambda^{\rm opt},\xi^{\rm opt})$ into \eqref{opt_r_g} and \eqref{optimal_Jg}.
\end{algorithmic}
\end{algorithm}

\subsubsection{Subproblem with respect to $\br$ and $\bmu$}
For given $\bJ$, the subproblem for the optimization of $\br$ and $\bmu$ is simplified into:
\begin{subequations}\label{subproblem_r_mu}
\begin{align}
  \max_{\br,\bmu} \quad & \sum_{g=1}^{G} \log_2\left(1+r_g\right)\\
  \st \quad & r_g \leq y(\mu_{g,k},J_g(\bs)),\forall g \in \cG, \forall k \in \cK_g \label{r_g_right_hand},\\
   \quad & r_g \geq 2^{\bar{R}_g}-1,\forall g \in \cG.
\end{align}
\end{subequations}
As discussed in \lemmref{mu_k_opt},  the optimal $\bmu^{\rm opt} \triangleq \{\mu^{\rm opt}_{g,k}\}_{\forall g \in \cG, \forall k \in \cK_g}$ to problem \eqref{subproblem_r_mu} is given by
\begin{align}\label{mu_update}
  \mu_{g,k}^{\rm opt} = \frac{\int_{\mathcal{S}_{\mathrm{T}}} h_{g,k}(\bs) J_g(\bs) d \mathbf{s}}{\sum_{i \in \cG \backslash \{g\}} \left|\int_{\mathcal{S}_{\mathrm{T}}} h_{g,k}(\bs) J_i(\bs) d \mathbf{s}\right|^2+\sigma_{g,k}^2}.
\end{align}
Substituting \eqref{mu_update} into $y(\mu_{g,k},J_g(\bs))$,   the optimal $\br^{\rm opt}$  is given by
\begin{align}\label{r_update}
  r_g^{\rm opt} = \min_{\forall k \in \cK_g} \frac{\left|\int_{\mathcal{S}_{\mathrm{T}}} h_{g,k}(\bs) J_g(\bs) d \mathbf{s}\right|^2}{\sum_{i \in \cG \backslash \{g\}} \left|\int_{\mathcal{S}_{\mathrm{T}}} h_{g,k}(\bs) J_i(\bs) d \mathbf{s}\right|^2+\sigma_{g,k}^2}.
\end{align}

\subsection{Convergence and Complexity Analysis}\label{subsec:complexity and convergence}
Based on the previous two subproblems, we propose a CoV-based BCD algorithm for solving problem \eqref{Before_BCD_problem}, which is summarized in \alref{alg:BCD}.
\begin{algorithm}[htbp]
\caption{Proposed CoV-based BCD Algorithm for Solving Problem \eqref{Before_BCD_problem}}
\label{alg:BCD}
\begin{algorithmic}[1]
\Require Feasible $\{\bJ^{(0)}(\bs),\br^{(0)},\bmu^{(0}\}$; iterative index $n=0$; threshold ${\varepsilon}$.
\State Calculate the objective function $f^{(n)} = R(\br) - \eta^{(\ell)}P(\bJ)$ in \eqref{Dinkel_obj}.
\Repeat
\State Solve subproblem \eqref{subproblem_J_r} with given $\bmu^{(n)}$ via \alref{alg:ellipsoid}, and denote the optimized solution as $\{\bJ^{(n+1)},\br^{(n+0.5)}\}$;
\State Solve subproblem \eqref{subproblem_r_mu} with given $\bJ^{(n+1)}$ using \eqref{mu_update} and \eqref{r_update}, and denote the optimized solution as $\{\bmu^{(n+1)},\br^{(n+1)}\}$;
\State Calculate the objective function $f^{(n+1)} = R(\br^{(n+1)}) - \eta^{(\ell)}P(\bJ^{(n+1)})$ in \eqref{Dinkel_obj};
\State Update $n = n+1$;
\Until{$\left|  f^{(n+1)} -  f^{(n)} \right|\le \varepsilon$}.
\Ensure The optimized source current pattern function $\{\bJ^{(n)}, \br^{(n)}, \bmu^{(n)}\}$.
\end{algorithmic}
\end{algorithm}
By using the proposed CoV-based BCD algorithm, we iteratively update $\{\bJ,\br\}$ and $\{\br,\bmu\}$ until the objective function in \eqref{Din_ori_obj} converges. 
We first demonstrate the complexity of the proposed CoV-based BCD algorithm for solving problem \eqref{Before_BCD_problem}. 
The computational complexity of the proposed BCD algorithm mainly  comes from solving the $\{\br,\bJ\}$ subproblem. In each iteration of the BCD algorithm, the computational complexity consists of the following three aspects:  \emph{\romannumeral1)} the complexity of calculating $\bQ$ in \eqref{cal_Q} is $\cO(M^2(K-K_g)^2)$ when using the $M$-point Gauss-Legendre quadrature; \emph{\romannumeral2)}  the complexity of matrix inversion involved in calculating $c_{j,i}$ in \eqref{cal_c_ji} is $\cO((K-K_g)^3)$; \emph{\romannumeral3)}  the complexity of the ellipsoid method used when solving the Lagrange dual problem is $\cO(K^4/\varepsilon)$\cite{ellipsoid}, where $\varepsilon$ is the threshold. As a result, the complexity of the BCD algorithm is $\cO(I_o(K^4/\varepsilon+(K-K_g)^3+M^2(K-K_g)^2))$, where $I_o$ is the number of BCD iterations.
Then, we analyze the convergence behavior of the proposed CoV-based BCD algorithm. 
Denote the objective function in \eqref{Din_ori_obj} as $f(r_g,J_g,\mu_{g,k})$. The proposed algorithm satisfies the following relationship during the iteration:
\begin{subequations}
  \begin{align}
 f(\bJ^{n},\br^{n},\bmu^{n})&\overset{\text{(a)}}{\leq} f(\bJ^{n+1},\br^{n+0.5},\bmu^{n}),\\
    &\overset{\text{(b)}}{\leq} f(\bJ^{n+1},\br^{n+1},\bmu^{n+1}),
  \end{align}
\end{subequations}
where (a) holds because when $\bmu$ is fixed, the $\{\bJ,\br\}$ subproblem in \eqref{subproblem_J_r} becomes jointly convex with respect to $\bJ$ and $\br$, and the subproblem is optimally solved; (b) holds because the $\{\br,\bmu\}$ subproblem is also a convex problem, and has closed-form solutions.

Based on the proposed CoV-based BCD algorithm, the  Dinkelbach's algorithm for solving problem \eqref{Dinkel_problem} is summarized in \alref{alg:Dinkelbach}.  Let $I_D$ denote the total number of iterations of the Dinkelbach's algorithm. The complexity of it is $\cO(I_D I_o(K^4+(K-K_g)^3+M^2(K-K_g)^2))$ \cite{zappone2015energy,9309152}.
\begin{algorithm}[htbp]
\caption{Dinkelbach's Algorithm for Solving Problem \eqref{Dinkel_problem}}
\label{alg:Dinkelbach}
\begin{algorithmic}[1]
\Require Threshold ${\varepsilon}$, and set iteration index $\ell = 0$.
\State Initialize $\eta^{(0)}$.
\Repeat
\State Solve problem \eqref{Dinkel_problem} via \alref{alg:BCD};
\State Set $\ell=\ell+1$;
\State Update ${\eta}^{(\ell)}$ by \eqref{Dinkel_update};
\Until{$\left|  {\eta}^{(\ell)} -  {\eta}^{(\ell-1)} \right|\le \varepsilon$}.
\Ensure The optimized CAPA beamformer $\bJ^{(\ell)}$.
\end{algorithmic}
\end{algorithm}

\section{Low-complexity ZF-based BCD Algorithm}\label{Sec:low complexity}
Note that the proposed CoV-based BCD algorithm might have high computational complexity due to the iterations required by the CAPA beamformer design.
In this section, we will propose a low-complexity ZF-based CAPA algorithm within the framework of Dinkelbach's method.
After designing the ZF-based CAPA beamformer, the original optimization problem \eqref{Dinkel_problem} is transformed into a simplified power allocation problem. 

\subsection{ZF-based Multicast Beamformer Design}
The key objective is that the  CAPA multicast beamformer design should maximize the intra-group worst-case performance while mitigating the inter-group interference.
This  design principle motivates us to utilize the ZF beamformer, since it is well-suited for eliminating the inter-user interference. However, with each group consisting of more than one user, it is hard to mathematically express the inter-group interference.
To address this issue, we develop a channel correlation-based user selection strategy and only utilize the selected users' channel information, rather than that of all users, to design the ZF beamformer, thereby reducing the computational complexity. 
In the following, we first introduce the user selection scheme based on the channel correlation and then derive the closed-form CAPA beamformer with the ZF principle.
\subsubsection{Channel Correlation-based User Selection}
Consider $|\cG| = G$ groups and that each group $g\in \cG$ has $|\cK_g| = K_g$ users.   Let $l_{i,j}\triangleq\langle h_i,h_j \rangle$ denote the channel correlation coefficients between user $i \in \cK$ and $j \in \cK$. Therefore, the sum of intra-group channel correlations and the sum of inter-group correlations of user $i \in \cK_g$ are given by
$O_i \triangleq \sum_{j\in \cK_g \backslash \{i\}} l_{i,j}$ and $\tilde{O}_{i} \triangleq \sum_{j \in \cK \backslash \cK_g} l_{i,j}$, respectively. 

To maximize the multi-group multicast performance, the user channel selected for each group should have a good representation of the other remaining users' channels, i.e., it should have the highest channel correlations with the other users' channels within the group. Motivated by this, the selected user channel $h_g(\bs)$ for group $g,\forall g\in \cG$, is firstly determined based on the sum of the intra-user channel correlations as follows:
\begin{align}\label{user_select_Kg3}
  h_g(\bs) = \arg \max_{\{h_{i}\}_{\forall i \in \cK_g}} O_{i}(h_i(\bs)),\quad \forall g\in \cG.
\end{align}
By doing so, the remaining users in each group would also achieve a relatively strong performance using the designed multicast CAPA beamformer based only on the selected user channels.  

Moreover, if there are more than one user selected from \eqref{user_select_Kg3}, i.e., the sum of intra-user channel correlations being identical, the representative user is further selected based on the sum of inter-user channel correlations with other groups as follows: 
\begin{align}
  h_g(\bs) = \arg \min_{\{h_j(\bs)\}_{\forall j \in \cK_g^*}} \tilde{O}_{j}(h_j(\bs)),\quad \forall g\in \cG,
\end{align}
where $\cK_g^*$ is the set of users having the same  sum of intra-group channel correlations. In this case, we can ensure that the inter-group interference can be well mitigated.

\subsubsection{Closed-Form ZF Beamformer}
Denote the set of representative users as $\cK_r$ and their channels set as $\cH_r\triangleq \{h_1,h_2,\dots,h_G\}$, where $(g,i)\in\cK_r$ and $h_{g,i}\triangleq h_g$.
Building upon these selected users' channels, we design the ZF-based beamformer to transform the problem into a power allocation problem among different groups, further reducing the complexity of the solution process.

With the ZF beamformer, the current density function $\bJ^{\rm ZF}\triangleq \{J_g^{\rm ZF}(\bs)\}_{\forall g \in\cG}$ should satisfy
\begin{align}\label{ZF_condition}
  \int_{\mathcal{S}_{\mathrm{T}}} h_i(\bs)J_g^{\rm ZF}(\bs) d \mathbf{s} = 0,\forall h_i \in \cH_r \backslash\{h_g\}.
\end{align}
In this case, the structure of $\bJ^{\rm ZF}$ is given in the following proposition:
\begin{proposition}\label{prop_ZF_J}
The ZF structure of $J_g^{\rm ZF}(\bs)$ is
\begin{subequations}\label{ZF_Jg}
\begin{align}
    J_g^{\rm ZF}(\bs) =& \sqrt{\rho_g} \hat{J}_{g}(\bs),\\
    \hat{J}_{g}(\bs) =& \frac{Z_g(\bs)}{\sqrt{\int_{\cS_T} |Z_g(\bs)|^2 d \bs}},\\
    Z_g(\bs) =& \sum_{j\in \cK_r}\beta_{g,j} h_{j}^*(\bs) ,
 \end{align}
\end{subequations}
 where $\rho_g$ is the power scaling factor to make the power transmitted to the $g$-th user $\rho_g$; $\bbeta_{g}\triangleq[\beta_{g,1},\beta_{g,2},\dots,\beta_{g,G}]^T$ is the $g$-th column of $\bH_{\rm ot}^{-1}$, where the $(g,j)$-th element of $\bH_{\rm ot}$ is defined as follows:
 \begin{align}
  [\bH_{\rm ot}]_{g,j} = \langle h_g,h_j \rangle,\quad \forall g,j \in \cK_r.
 \end{align}
\end{proposition}
\begin{proof}
  Substituting \eqref{ZF_Jg} into \eqref{ZF_condition}, we have
 \begin{subequations}
    \begin{align}
      &\int_{\mathcal{S}_{\mathrm{T}}} h_i(\bs) J_g(\bs) d \mathbf{s} \notag \\
      &=  \frac{\sqrt{\rho_g}}{\sqrt{\int_{\cS_T} |Z_g(\bs)|^2 d \bs}}\sum_{j\in \cG} \beta_{g,j} \int_{\mathcal{S}_{\mathrm{T}}} h_i(\bs)h_{j}^*(\bs) d \mathbf{s} \\
     &= \frac{\sqrt{\rho_g}}{\sqrt{\int_{\cS_T} |Z_g(\bs)|^2 d \bs}} [\bH_{\rm ot}]_{i,:}  \bbeta_g ,
    \end{align}
 \end{subequations}
  where $[\bH_{\rm ot}]_{i,:}$ is the $i$-th row of $\bH_{\rm ot}$. When $i \neq g$,  $[\bH_{\rm ot}]_{i,:}  \bbeta_g = 0$, and when $i = g$, $[\bH_{\rm ot}]_{i,:}  \bbeta_g = 1$, which completes the proof.
\end{proof}

By utilizing the ZF beamformer, the SINR of user $(g,k)$ will become
\begin{align}
  \gamma_{g,k} = \frac{\frac{\rho_g}{P_g} |c_{g,k}^g|^2}{\sum_{i \in \cG \backslash \{g\}}\frac{\rho_i}{P_i} |c_{g,k}^i|^2+\sigma_{g,k}^2},
\end{align}
where $c_{g,k}^i \triangleq \int_{\mathcal{S}_{\mathrm{T}}} h_{g,k}(\bs) Z_i(\bs) d\bs, \forall i,g \in \cG,\forall k \in\cK_g$ and
 $P_{g}\triangleq \int_{\mathcal{S}_{\mathrm{T}}} |Z_g(\bs)|^2 d \bs$.
Specifically, $c_{g,k}^i = 0$ when user $(g,k) \in \cK_r$ and $i\neq g$.

Similarly to \lemmref{mu_k_opt},  we introduce another set of auxiliary variables $\tilde{\bmu} \triangleq \{\tilde{\mu}_{g,k}\}_{g \in \cG,k \in \cK_g}$ and transform the minimum multicast SE constraint into the following one:
\begin{align}
&y(\tilde{\mu}_{g,k},\rho_g)\notag \\
& \triangleq 2\tilde{\mu}_{g,k}\sqrt{\frac{\rho_g}{P_g}}|c_{g,k}^g | - \tilde{\mu}_{g,k}^2 \left(\!\sum_{i \in \cG \backslash \{g\}}\!\frac{\rho_i}{P_i}|c_{g,k}^i |^2 + \sigma_{g,k}^2\right).
\end{align}
The reformulated problem is therefore given by
\begin{subequations}\label{ZF_orig_problem}
  \begin{align}
  \max_{\{\br, \brho,\tilde{\bmu}\}} \quad & \sum_{g=1}^{G}\log_2 (1+r_g)- \eta^{(\ell)}\sum_{g=1}^{G}\rho_g\label{obj_ZF-based}\\
     \st \quad &  r_g \leq y(\tilde{\mu}_{g,k},\rho_g),\forall k \in \cK_g, g \in \cG, \label{reformulated_SINR_constraint_ZF}\\
     & r_g \geq 2^{\bar{R}_g}-1,\forall g \in \cG,\label{minimum_rate_constraint_ZF}\\
    &\sum_{g=1}^{G}\rho_g \leq P_t,\label{power_constraint_ZF}\\
    &\rho_g \geq 0,\forall g \in \cG\label{power_geq_zero_constraint_ZF}.
  \end{align}
\end{subequations}
The original problem \eqref{Dinkel_problem} has been simplified to a power allocation problem across different groups, which will be addressed by employing the BCD method.

\subsection{ZF-based BCD algorithm}
Similarly, problem \eqref{ZF_orig_problem} can be decomposed into subproblems with respect to $\{\br,\brho\}$ and $\{\br,\tilde{\bmu}\}$, where $\brho \triangleq \{\rho_g\}_{g=1}^G$.
Since the  procedures of solving the problem are similar to that in \secref{Sec:CoV-based}, the details are briefly explained as follows:

\subsubsection{Subprobelm with respect to $\br$ and $\brho$}For given $\tilde{\bmu}$, 
we have the following subproblem:
\begin{subequations}\label{subproblem_ZF_r_rho}
  \begin{align}
     \max_{\{\br, \brho\}} \quad & \sum_{g=1}^{G}\log_2 (1+r_g)- \eta^{(\ell)}\sum_{g=1}^{G}\rho_g\\
     \st \quad  & \eqref{reformulated_SINR_constraint_ZF}-\eqref{power_geq_zero_constraint_ZF}.
  \end{align}
\end{subequations}
It can be observed that problem \eqref{subproblem_ZF_r_rho} is convex, which can be efficiently solved using existing tools, such as CVX \cite{CVXtool}.

\subsubsection{Subprobelm with respect to $\br$ and $\tilde{\bmu}$} For given $\brho$, 
we have the following simplified subproblem,
\begin{subequations}\label{subproblem_ZF_r_mu}
  \begin{align}
    \max_{\{r_g,\tilde{\mu}_{g,k}\}} \quad & \sum_{g=1}^{G}\log_2 (1+r_g)\\
     \st \quad &  \eqref{reformulated_SINR_constraint_ZF},\eqref{minimum_rate_constraint_ZF}.
  \end{align}
\end{subequations}
The optimal $\tilde{\bmu}^{\rm opt}\triangleq \{\tilde{\mu}^{\rm opt}_{g,k}\}_{\forall g \in \cG,\forall k \in \cK_g}$ is given by similarly utilizing \lemmref{mu_k_opt}, which is given by
\begin{align}\label{ZF_mu_update}
  \tilde{\mu}_{g,k}^{\rm opt} = \frac{\sqrt{\frac{\rho_g}{P_g}}|c_{g,k}^g|}{\sum_{i \in \cG \backslash \{g\}}\frac{\rho_i}{P_i}|c_{g,k}^i |^2 + \sigma_{g,k}^2},\forall g \in\cG,\forall k \in \cK_g,
\end{align}
while the optimal $\br$ is therefore given by
\begin{align}\label{ZF_r_update}
  r_g^{\rm opt} = \min_{k \in \cK_g} \frac{\frac{\rho_g}{P_g}|c_{g,k}^g|^2}{\sum_{i \in \cG \backslash \{g\}}\frac{\rho_i}{P_i}|c_{g,k}^i |^2 + \sigma_{g,k}^2},\forall g\in\cG.
\end{align}

\subsection{Convergence and Complexity Analysis}
\begin{algorithm}[t]
\caption{Proposed ZF-based BCD Algorithm}
\label{alg:ZF-based algorithm}
\begin{algorithmic}[1]
\Require Feasible $\{\brho^{(0)}(\bs),\br^{(0)},\tilde{\bmu}^{(0}\}$; iterative index $m=0$; threshold ${\varepsilon}$.
\State Select $G$ representative users according to \eqref{user_select_Kg3}.
\State Calculate the objective function $f^{(m)} = \sum_{g=1}^{G}\log_2 (1+r_g^{(m)})- \eta^{(\ell)}\sum_{g=1}^{G}\rho_g^{(m)}$ in \eqref{obj_ZF-based}.
\Repeat
\State Solve subproblem \eqref{subproblem_ZF_r_rho} with given $\tilde{\bmu}^{(m)}$, and denote the optimized solution as $\{\brho^{(m+1)},\br^{(m+0.5)}\}$;
\State Solve subproblem \eqref{subproblem_ZF_r_mu} with given $\brho^{(m+1)}$ using \eqref{ZF_mu_update} and \eqref{ZF_r_update}, and denote the optimized solution as $\{\bmu^{(m+1)},\br^{(m+1)}\}$;
\State Calculate the objective function $f^{(m+1)} = \sum_{g=1}^{G}\log_2 (1+r_g^{(m+1)})- \eta^{(\ell)}\sum_{g=1}^{G}\rho_g^{(m+1)}$ in \eqref{obj_ZF-based};
\State Update $m = m+1$;
\Until{$\left|  f^{(m+1)} -  f^{(m)} \right|\le \varepsilon$}.
\Ensure The optimized source current pattern function $\{\brho^{(m)}, \br^{(m)}, \tilde{\bmu}^{(m)}\}$.
\end{algorithmic}
\end{algorithm}
By using the proposed ZF-based BCD algorithm, we iteratively update $\{\br,\brho\}$ and $\{\br,\tilde{\bmu}\}$ until the objective function of \eqref{obj_ZF-based} converges.
 The ZF-based BCD algorithm as well as the user selection has been summarized in \alref{alg:ZF-based algorithm}.
The convergence behaviour of this ZF-based BCD algorithm can be similarly analyzed as that in \subsecref{subsec:complexity and convergence}, and thus it is omitted here for brevity.

The computational complexity of the ZF-based BCD algorithm mainly lies in the following two aspects. First, the complexity of the one-time calculation of $\bH_{\rm ot}^{-1}$  is $\cO(M^2G^2+G^3)$ when using the $M$-point Gauss-Legendre quadrature.  Second, the complexity of using the interior-point method to solve the convex subproblem \eqref{subproblem_ZF_r_rho}  is $\cO((2G)^{3.5})$ \cite{boyd2004convex}. Thus, the complexity of the ZF-based BCD algorithm is $\cO(I_o(2G)^{3.5}+G^3+M^2G^2)$, where $I_o$ is the iterations of the BCD algorithm.  Based on this,  the complexity of the Dinkelbach's algorithm is $\cO(I_D(I_o(2G)^{3.5}+G^3+M^2G^2) )$, where $I_D$ is the number of iterations in Dinkelbach's algorithm.
It can be observed that unlike the proposed CoV-based BCD algorithm, the ZF-based BCD algorithm does not require iterations when obtaining the optimized power $\brho$, since the structure of the beamformer is designed based on the ZF criterion. Thus, there is no need to resort to the CoV theory and ellipsoid method, thereby highly reducing the computational complexity.

\section{Simulation Results}\label{Sec:simuations}
In this section, we demonstrate the effectiveness of the proposed CAPA-based multi-group multicast scheme.  
\subsection{Simulation Setup and Baselines}
We consider a 3D simulation setup, where the BS is equipped with a CAPA centered at the origin of the coordinate system, and
\begin{align}
  \cS_T: \{[s_x,s_y,0]^T\big| |s_x|\leq L_x/2, |s_y|\leq L_y/2\},
\end{align}
with $L_x = L_y = 0.5$ m.
The users' positions are randomly generated according to the following parameters: the distance between different groups is randomly from 1 m to 5 m; users' spread radius, which refers to the radius of the minimum enclosing circle encompassing all users within the same group, is 1 m; the group centers $[r_x,r_y,r_z]^T$ are randomly generated within the following space:
\begin{align}
\cV:\{[r_x,r_y,r_z]^T\big|-5 \leq r_x,r_y \leq 5, 15 \leq r_z \leq 30\}.
\end{align}
Moreover, the signal frequency is 2.4 GHz and the free-space impedance is $\eta = 120 \pi$ $\Omega$. The transmit power and the noise power at the receiver are $P_t = 1000$ mA$^2$ and $5.6\times 10^{-3}$ V$^2$/m$^2$ \cite{sanguinettiWavenumberDivisionMultiplexingLineofSight2023}, respectively. The numbers of groups and users within each group are specifically clarified in the following subsections. 
 The polarizations of all users are set to $\hat{\bu}_x = \hat{\bu}_y = [0,1,0]^T$.
The integrals in the paper are computed by using the Gauss-Legendre quadrature, and the number of samples are 20. The thresholds for the BCD algorithm and the Dinkelbach's algorithm are both $10^{-4}$. The results in Figs. 4-7 are obtained by averaging over 50 channel realizations. Specifically, we first randomly generate 50 user distributions, and then obtain the corresponding channel realizations.

For performance comparison, we mainly consider the following SPDA  scheme, where the continuous surface $\cS_T$  is equipped with discrete antennas with $\lambda/2$ spacing. The location of the $(n_x,n_y)$th antenna is 
\begin{align}
  \bs_{n_x,n_y} = \left[(n_x-1)d-\frac{L_x}{2},(n_y-1)d-\frac{L_y}{2},0\right]^T,
\end{align}
while the number of the transmit antenna is $N = N_x \times N_y$, with $N_x = \lceil L_x/d \rceil$ and $N_y = \lceil L_y/d \rceil$. 

The $(n_x,n_y)$th element of the channel vector $\bh_{g,k}$ is
\begin{align}
  [\bh_{g,k}]_{n_x,n_y} = \sqrt{\frac{\lambda^2}{4\pi}} \hat{\bu}_k^T \bG(\br_{k},\bs_{n_x,n_y})\hat{\bu}_y,
\end{align}
where $\sqrt{\lambda^2/(4\pi)}$ is the effective area of the antenna.

Given the above discrete channels, it becomes a conventional MIMO beamforming optimization problem for maximizing the system EE. The ZF method can also be exploited for it to reduce the beamforming complexity.
Since the procedures to design the beamformer are similar, they are omitted here for the sake of brevity.

\subsection{Convergence Performance}
\begin{figure}[htbp]
    \captionsetup{font={small}}
    \centering
    \includegraphics[width = 0.45\textwidth]{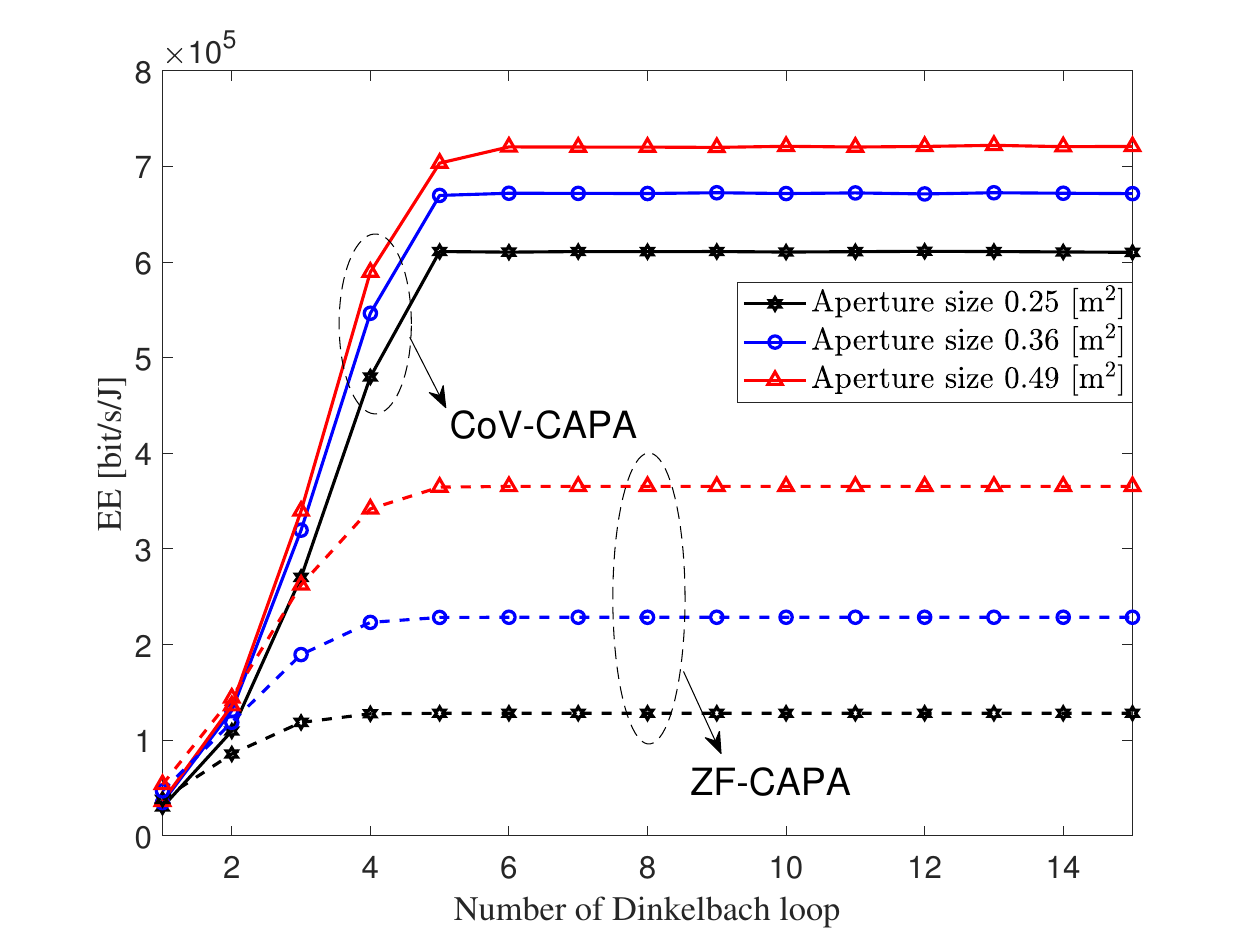}
    \caption{Convergence of the proposed CAPA algorithm, with $G = 3$ and $K_g = 3$.}
    \label{fig_convergence_aperture_size}
\end{figure}
In \figref{fig_convergence_aperture_size}, we investigate the convergence behavior of the proposed CAPA and ZF-CAPA algorithms. Generally, we consider three different aperture sizes at the BS, with  $G = 3$ and $K_g  = 3$.  The initial  $\bJ$, $\br$, and $\bmu$ for the proposed CAPA algorithm are obtained with the following method.
\begin{itemize}
  \item CAPA beamformer $\bJ^{(0)}$: The structure of the initial $\bJ$ is given by the ZF beamformer, which is given in \propref{prop_ZF_J}. Then, we randomly generate the power allocated to each group, making sure that the total transmit power budget is satisfied.
  \item  $\br^{(0)}$ and $\bmu^{(0)}$: Utilize the initialized $\bJ^{(0)}$ and \eqref{mu_update} to initialize $\bmu^{(0)}$, while
  utilize the initialized $\bJ^{(0)}$ and \eqref{r_update} to initialize $\br^{(0)}$.
\end{itemize}
The presented results are obtained for one random channel realization.
It can be observed that the convergence behavior of the proposed CAPA algorithm is almost the same as that of the proposed ZF-CAPA algorithm, both of which converge quickly within 10 Dinkelbach loops. Meanwhile,  the complexity of both algorithms does not scale with the aperture size $S_T$ generally. This makes the CAPA configuration quite appealing for practical implementation  since the complexity in SPDA increases with the number of antennas, e.g., in massive MIMO.

\subsection{Impact of Aperture Size}\label{subsection:impact of aperture}
\begin{figure}[htbp]
    \captionsetup{font={small}}
    \centering
    \includegraphics[width = 0.45\textwidth]{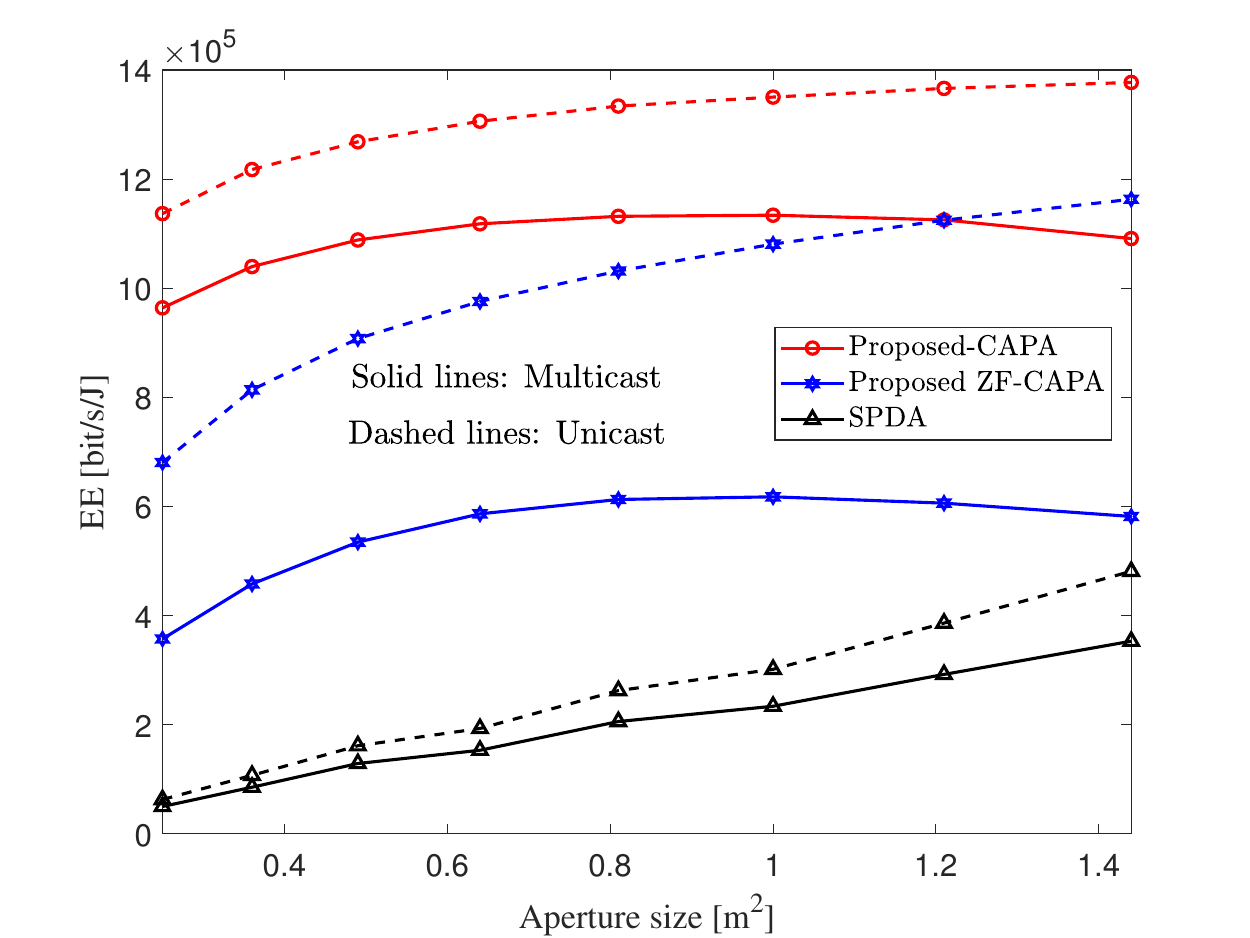}
    \caption{EE versus the aperture size, with $G = 3$ and $K_g = 3$ in multicast and $G = 3$ and $K_g = 1$ in unicast.}
    \label{fig_impact_aperture_size}
\end{figure}
In \figref{fig_impact_aperture_size}, we investigate the impact of the aperture size $S_T$ on the EE of the CAPA-based multi-group multicast system, with $G = 3$ and $K_g = 3$. We consider the unicast scenario as a baseline here by setting $K_g = 1$. 
It can be observed that the EE obtained by CAPA is always higher than that obtained by SPDA when the aperture has a moderate size, e.g.,  when the aperture size $S_T = 1$ m$^2$, at which point the SPDA corresponds to a $9\times9$ antenna array.
However, different from the EE obtained by SPDA, which continues to grow with the increase of the aperture size $S_T$, the EE obtained by the proposed CAPA scheme and the ZF-CAPA scheme decrease whenever $S_T$ increases after a certain value  in multicast scenarios. 
Meanwhile, in unicast scenarios, the EE obtained by all the schemes is growing with the increase of $S_T$.
The reason behind this phenomenon can be explained as follows.
While the increase of $S_T$ does help to diminish the inter-user interference, it also diminishes the correlations of the channels among the users in the same group, i.e., the users within the same group may become orthogonal to each other if the aperture size $S_T$ continues to increase \cite{wang2025beamforming}. Thus, it becomes increasingly challenging to sustain the intra-group received signal strength for users, which  deteriorates the EE of the system.  
Note that the decrease of EE shown in the proposed CAPA scheme is slighter compared to that obtained in the ZF-CAPA scheme, as the proposed CAPA scheme takes into account the channel characteristics of the entire group of users during optimization, which shows the importance of the CAPA beamformer design.
As for the continuous growth of the EE in the SPDA scheme, this can be attributed to the fact that the DoFs of SPDA are always limited.  The increase in $S_T$ does not significantly reduce the correlation between different users' channels. 
Summarizing,  the obtained results suggest that a moderate $S_T$ is more favorable for the enhancement of the multi-group multicast system EE.

\subsection{Impact of User Spread Radius}
\begin{figure}[t]
\captionsetup{font={small}}
    \centering
    \includegraphics[width = 0.45\textwidth]{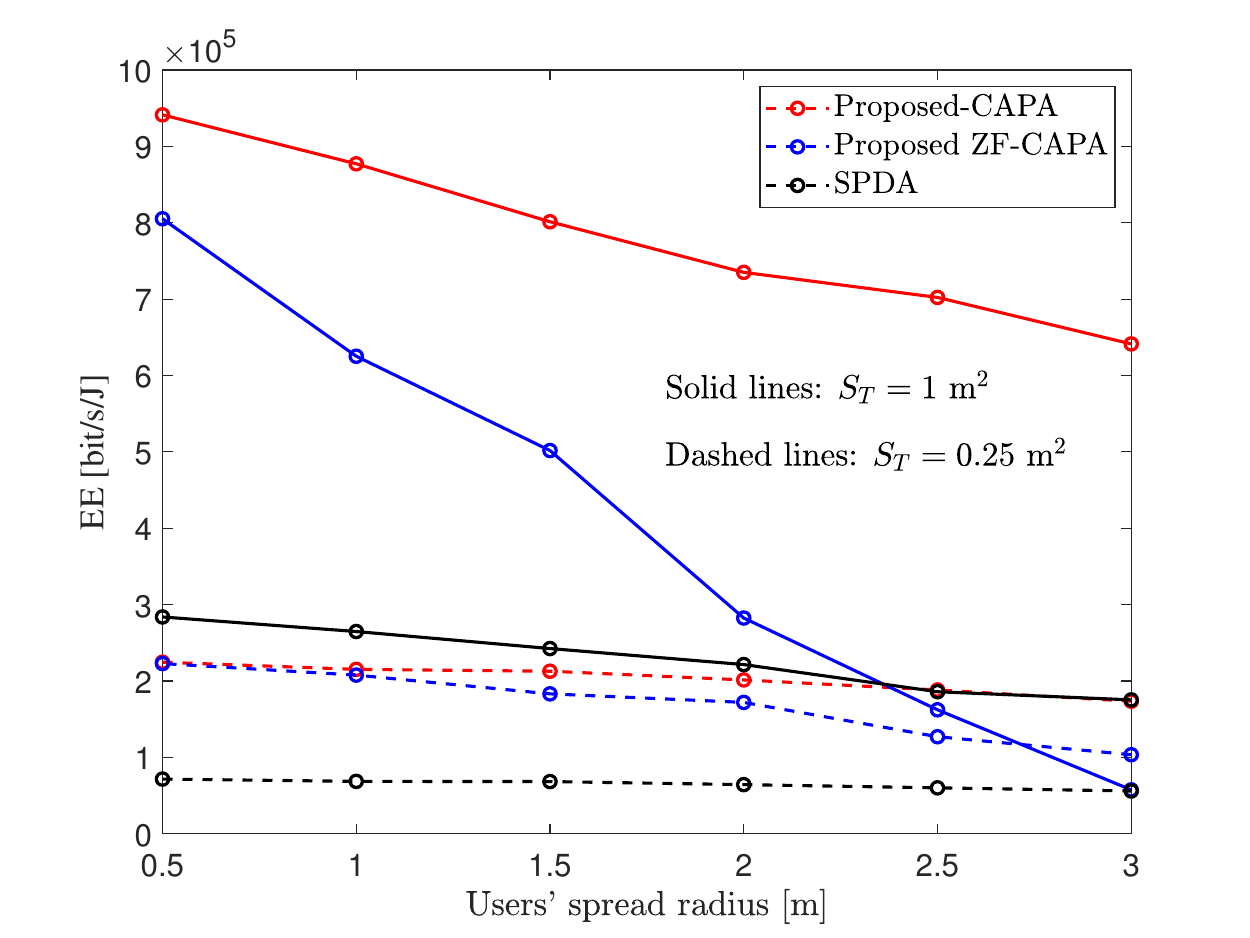}
    \caption{EE versus the users' spread radius, with $G = 3$ and $K_g = 2$.}
    \label{fig_impact_user_spread_radius}
\end{figure}
In \figref{fig_impact_user_spread_radius}, we investigate the impact of the users' spread radius on the CAPA-based multi-group multicast system EE, with $G = 3$ and $K_g = 2$.
When the aperture size $S_T = 0.25$ m$^2$, we can observe that the EE achieved by the proposed ZF-CAPA scheme can achieve near-optimal performance when the users' spread radius is relatively small.
However, as the users' spread radius increases, the EE obtained by the proposed ZF-CAPA scheme declines more significantly than the proposed CAPA scheme.
When $S_T = 1$ m$^2$, the EE gap between the proposed ZF-CAPA scheme and the CAPA scheme becomes much larger. Moreover, when the users' spread radius increases beyond a certain level, the EE achieved by ZF-CAPA can become even worse than that of SPDA. 
These are all due to the fact that the ZF-CAPA beamformer only uses one of the user's channel in each group and does not fully utilize the channel information of the entire group when designing the beamformer. As the users' spread radius increases, the channels of different users within a group tend to be orthogonal to each other.  This makes it difficult for a single user's channel to be representative of the entire group of users, which exactly illustrates the importance of utilizing the full user channel information to design the CAPA beamformer. This observation aligns with the earlier analysis in \subsecref{subsection:impact of aperture}.

\subsection{Impact of User Numbers}
\begin{figure}[t]
\captionsetup{font={small}}
    \centering
    \includegraphics[width = 0.45\textwidth]{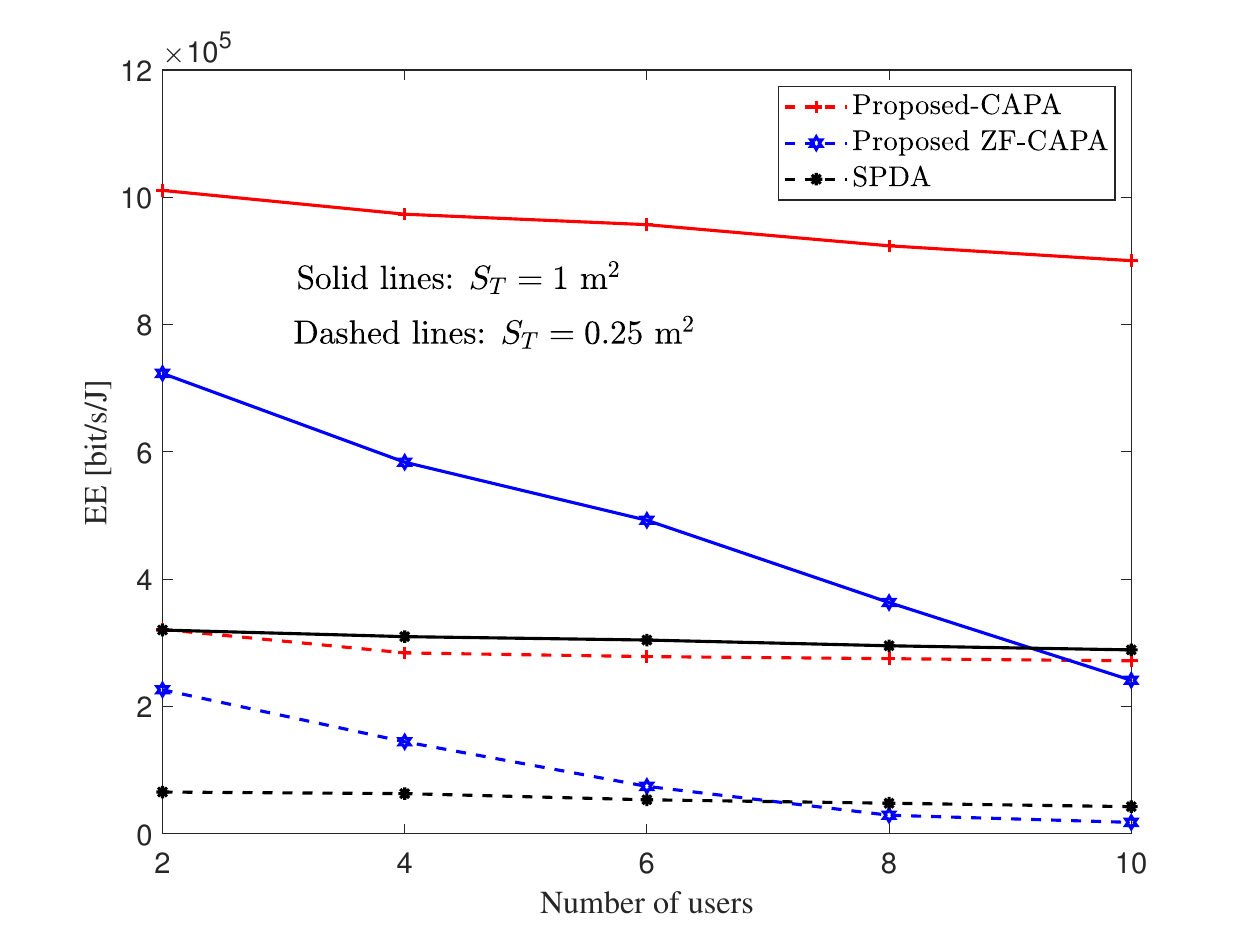}
    \caption{EE versus the number of users.}
    \label{fig_impact_users_number}
\end{figure}
In \figref{fig_impact_users_number}, we investigate the impact of the users' number on the CAPA-based multi-group multicast system  SE. We keep $K_g = 2$ and increase the number of users in the system by increasing $G$ from 1 to 5. First, it can be observed that when $S_T = 0.25$ m$^2$, the proposed ZF-CAPA scheme achieves a EE nearly identical to that of SPDA when the number of users reaches 6. However, as the number of users continues to increase, the EE achieved by ZF-CAPA becomes inferior to that of SPDA. This is because the ZF-CAPA scheme allocates more power to mitigate inter-user interference, which also reduces the effective intro-group signal strength. Meanwhile, it can be seen that the proposed CAPA achieves nearly the same EE with $S_T  = 0.25$ m$^2$ as SPDA does  with an aperture size of $S_T = 1$ m$^2$, which also shows a CAPA with moderate $S_T$ is sufficient to achieve a relatively good EE.
Moreover, it is worth noting that regardless of whether $S_T = 0.25$ m$^2$ or $S_T = 1$ m$^2$, the decline in the system EE achieved by the proposed CAPA scheme  is not significant when the number of users continues to increase. This demonstrates that a CAPA can accommodate more users while maintaining the same EE performance.

\subsection{Impact of Minimum Multicast Rate Constraint}
\begin{figure}[t]
\captionsetup{font={small}}
    \centering
    \includegraphics[width = 0.45\textwidth]{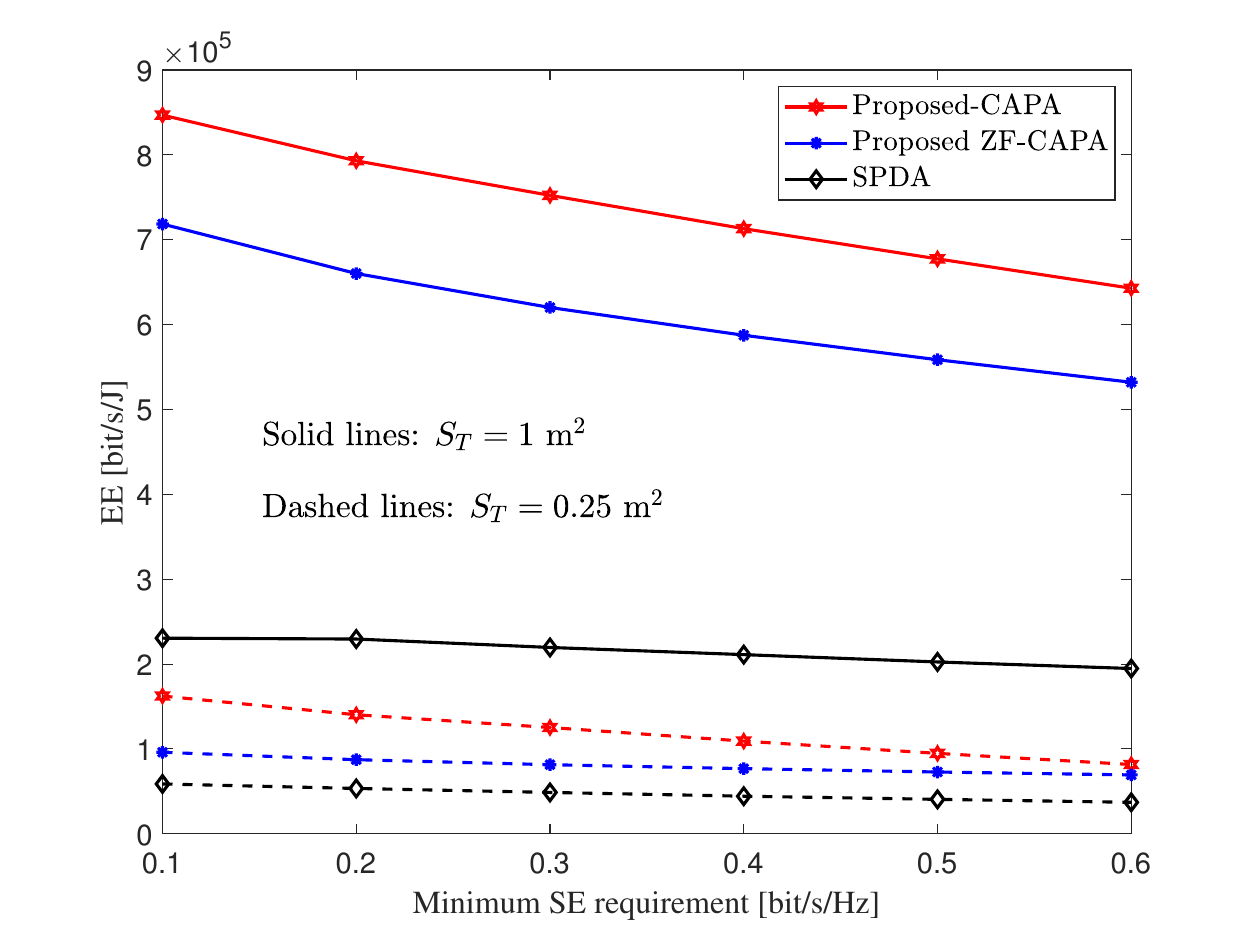}
    \caption{EE versus the minimum SE requirement, with $G=3$ and $K_g = 2$.}
    \label{fig_impact_minimum_rate}
\end{figure}
In \figref{fig_impact_minimum_rate}, we investigate the impact of the minimum SE constraint on the CAPA-based multi-group multicast system EE, with $G=3$ and $K_g = 2$. It can be observed that there is a decline in the EE achieved by all schemes as the minimum SE constraint increases. This is because, as the system minimum SE requirement increases, more power is needed to meet this demand, which leads to a decrease in the EE.
Nevertheless, the proposed CAPA scheme and the ZF-CAPA scheme consistently outperform the SPDA scheme in terms of EE, underscoring the advantage of CAPA in enhancing the system performance, and this advantage becomes more pronounced when $S_T$ is 1 m$^2$.

\section{Conclusions}\label{Sec:conclusions}
In this paper, a CAPA-based multi-group multicast communication system was studied. 
An EE maximization problem with the integral-based beamforming design was formulated, subject to both the minimum multicast SE constraint of each group and the transmit power constraint.  Within Dinkelbach's framework, an efficient BCD-based algorithm was developed, where the CAPA beamformer was obtained by solving the relevant subproblems with the help of Lagrangian dual method and CoV theory. Our design strategy showed that the optimized CAPA beamformer is the combination of all users' channels. To further reduce the complexity of the proposed CAPA algorithm, a ZF-based CAPA algorithm was proposed, which combined the user selection algorithm and ZF beamformer design.
Finally, our simulation results verified that the proposed CAPA and ZF-based CAPA algorithms can achieve higher  EE compared to SPDA.
However, large aperture sizes and users' spread radius may undermine the achieved EE, indicating that a  CAPA with a moderate aperture size is sufficient in practice. This motivates future research on the optimal aperture size design or surface selection for CAPA-based multicast communications.

\bibliographystyle{IEEEtran.bst}
\bibliography{Refabrv_20180802,ref_EE.bib}

\begin{thebibliography}{10}
\providecommand{\url}[1]{#1}
\csname url@samestyle\endcsname
\providecommand{\newblock}{\relax}
\providecommand{\bibinfo}[2]{#2}
\providecommand{\BIBentrySTDinterwordspacing}{\spaceskip=0pt\relax}
\providecommand{\BIBentryALTinterwordstretchfactor}{4}
\providecommand{\BIBentryALTinterwordspacing}{\spaceskip=\fontdimen2\font plus
\BIBentryALTinterwordstretchfactor\fontdimen3\font minus
  \fontdimen4\font\relax}
\providecommand{\BIBforeignlanguage}[2]{{%
\expandafter\ifx\csname l@#1\endcsname\relax
\typeout{** WARNING: IEEEtran.bst: No hyphenation pattern has been}%
\typeout{** loaded for the language `#1'. Using the pattern for}%
\typeout{** the default language instead.}%
\else
\language=\csname l@#1\endcsname
\fi
#2}}
\providecommand{\BIBdecl}{\relax}
\BIBdecl

\bibitem{10684260}
M.~Mohammadi, Z.~Mobini, H.~Q. Ngo, and M.~Matthaiou, ``Ten years of research
  advances in full-duplex massive {MIMO},'' \emph{{IEEE} Trans. Wireless
  Commun.}, vol.~73, no.~3, pp. 1756--1786, Mar. 2025.

\bibitem{zhang2020prospective}
J.~Zhang \emph{et~al.}, ``Prospective multiple antenna technologies for beyond
  {5G},'' \emph{{IEEE} J. Sel. Areas Commun.}, vol.~38, no.~8, pp. 1637--1660,
  Aug. 2020.

\bibitem{jin2024gdm4mmimo}
Z.~Jin \emph{et~al.}, ``{GDM4MMIMO}: {Generative} diffusion models for massive
  {MIMO} communications,'' \emph{arXiv preprint arXiv:2412.18281}, 2024.

\bibitem{10117500}
Y.~Han, S.~Jin, M.~Matthaiou, T.~Q.~S. Quek, and C.-K. Wen, ``Toward extra
  large-scale {MIMO}: {New} channel properties and low-cost designs,''
  \emph{{IEEE} Internet Things J.}, vol.~10, no.~16, pp. 14\,569--14\,594, Aug.
  2023.

\bibitem{xuNearFieldWidebandExtremely2022a}
J.~Xu \emph{et~al.}, ``Near-{{field wideband extremely large-scale MIMO
  transmission}} with {{holographic metasurface antennas}},'' \emph{{IEEE}
  Trans. Wireless Commun.}, vol.~23, no.~9, pp. 12\,054--12\,067, Sep. 2024.

\bibitem{wei2024electromagnetic}
L.~Wei \emph{et~al.}, ``Electromagnetic information theory for holographic
  {MIMO} communications,'' \emph{arXiv preprint arXiv:2405.10496}, 2024.

\bibitem{liu2024capa}
Y.~Liu \emph{et~al.}, ``{CAPA}: {Continuous-aperture} arrays for
  revolutionizing {6G} wireless communications,'' \emph{arXiv preprint
  arXiv:2412.00894}, 2024.

\bibitem{huang2020holographic}
C.~Huang \emph{et~al.}, ``Holographic {MIMO} surfaces for {6G} wireless
  networks: {Opportunities}, challenges, and trends,'' \emph{{IEEE} Wireless
  Commun.}, vol.~27, no.~5, pp. 118--125, Oct. 2020.

\bibitem{poonDegreesFreedomMultipleantenna2005}
A.~S.~Y. Poon, R.~Brodersen, and D.~N.~C. Tse, ``Degrees of freedom in
  multiple-antenna channels: {A} signal space approach,'' \emph{{IEEE} Trans.
  Inf. Theory}, vol.~51, no.~2, pp. 523--536, Feb. 2005.

\bibitem{poonDegreeFreedomGainUsing2011}
A.~S.~Y. Poon and D.~N.~C. Tse, ``Degree-of-freedom gain from using
  polarimetric antenna elements,'' \emph{{IEEE} Trans. Inf. Theory}, vol.~57,
  no.~9, pp. 5695--5709, Sep. 2011.

\bibitem{dardariCommunicatingLargeIntelligent2020}
D.~Dardari, ``Communicating with large intelligent surfaces: {{Fundamental
  limits}} and models,'' \emph{{IEEE} J. Sel. Areas Commun.}, vol.~38, no.~11,
  pp. 2526--2537, Nov. 2020.

\bibitem{9848802}
L.~Ding, E.~G. Str{\"o}m, and J.~Zhang, ``Degrees of freedom in {3D} linear
  large-scale antenna array communications—{A} spatial bandwidth approach,''
  \emph{{IEEE} J. Sel. Areas Commun.}, vol.~40, no.~10, pp. 2805--2822, Oct.
  2022.

\bibitem{sanguinettiWavenumberDivisionMultiplexingLineofSight2023}
L.~Sanguinetti, A.~A. D'Amico, and M.~Debbah, ``Wavenumber-division
  multiplexing in line-of-sight holographic {MIMO} communications,''
  \emph{{IEEE} Trans. Wireless Commun.}, vol.~22, no.~4, pp. 2186--2201, Apr.
  2023.

\bibitem{zhangPatternDivisionMultiplexingMultiUser2023}
Z.~Zhang and L.~Dai, ``Pattern-division multiplexing for multi-user
  continuous-aperture {{MIMO}},'' \emph{{IEEE} J. Sel. Areas Commun.}, vol.~41,
  no.~8, pp. 2350--2366, Aug. 2023.

\bibitem{qian2024spectral}
M.~Qian, L.~You, X.-G. Xia, and X.~Gao, ``On the spectral efficiency of
  multi-user holographic {MIMO} uplink transmission,'' \emph{{IEEE} Trans.
  Wireless Commun.}, vol.~23, no.~10, pp. 15\,421--15\,434, Oct. 2024.

\bibitem{wang2025beamforming}
Z.~Wang, C.~Ouyang, and Y.~Liu, ``Beamforming optimization for continuous
  aperture array {(CAPA)}-based communications,'' \emph{{IEEE} Trans. Wireless
  Commun.}, early access, doi:10.1109/TWC.2025.3545770, 2025.

\bibitem{wang2025optimalbeamformingmultiusercontinuous}
Z.~Wang, C.~Ouyang, and Y.~Liu, ``Optimal beamforming for multi-user continuous
  aperture array {(CAPA)} systems,'' \emph{{IEEE} Trans. Wireless Commun.},
  early access, doi:10.1109/TCOMM.2025.3554644, 2025.

\bibitem{securebeamformingCAPA2025}
M.~Sun, C.~Ouyang, Z.~Wang, S.~Wu, and Y.~Liu, ``Secure beamforming for
  continuous aperture array {(CAPA)} systems,'' \emph{arXiv preprint
  arXiv:2501.04924}, 2025.

\bibitem{jeonCapacityContinuousspaceElectromagnetic2018}
W.~Jeon and S.-Y. Chung, ``Capacity of continuous-space electromagnetic
  channels with lossy transceivers,'' \emph{{IEEE} Trans. Inf. Theory},
  vol.~64, no.~3, pp. 1977--1991, Mar. 2018.

\bibitem{mikkiShannonInformationCapacity2023}
S.~Mikki, ``The {Shannon} information capacity of an arbitrary radiating
  surface: {{An electromagnetic approach}},'' \emph{{IEEE} Trans. Antennas
  Propag.}, vol.~71, no.~3, pp. 2556--2570, Mar. 2023.

\bibitem{wan2023can}
Z.~Wan, J.~Zhu, and L.~Dai, ``Can continuous aperture {MIMO} obtain more mutual
  information than discrete {MIMO}?'' \emph{{IEEE} Commun. Lett.}, vol.~27,
  no.~12, pp. 3185--3189, Dec. 2023.

\bibitem{iacovelli2025multi}
C.~Iacovelli, G.~Iacovelli, and S.~Chatzinotas, ``Multi-user holographic
  communications via channel operator diagonalization,'' \emph{Authorea
  Preprints}, Mar. 2025.

\bibitem{9093950}
M.~Dong and Q.~Wang, ``Multi-group multicast beamforming: {Optimal} structure
  and efficient algorithms,'' \emph{{IEEE} Trans. Signal Process.}, vol.~68,
  pp. 3738--3753, Jun. 2020.

\bibitem{ellipsoid}
\BIBentryALTinterwordspacing
S.~Boyd, ``Ellipsoid method,'' {Stanford University}. [Online]. Available:
  \url{https://web.stanford.edu/class/ee364b/lectures/ ellipsoid method
  notes.pdf}
\BIBentrySTDinterwordspacing

\bibitem{zappone2015energy}
A.~Zappone and E.~Jorswieck, ``Energy efficiency in wireless networks via
  fractional programming theory,'' \emph{Found. Trends Commun. Inf. Theory},
  vol.~11, no. 3-4, pp. 185--396, 2015.

\bibitem{9309152}
L.~You \emph{et~al.}, ``Energy efficiency and spectral efficiency tradeoff in
  {RIS}-aided multiuser {MIMO} uplink transmission,'' \emph{{IEEE} Trans.
  Signal Process.}, vol.~69, pp. 1407--1421, Mar. 2021.

\bibitem{CVXtool}
\BIBentryALTinterwordspacing
M.~Grant and S.~Boyd, \emph{CVX: Matlab Software for Disciplined Convex
  Programming, Version 2.1}, Mar. 2014. [Online]. Available:
  \url{http://cvxr.com/cvx}
\BIBentrySTDinterwordspacing

\bibitem{boyd2004convex}
S.~P. Boyd and L.~Vandenberghe, \emph{Convex Optimization}.\hskip 1em plus
  0.5em minus 0.4em\relax Cambridge, U.K.: Cambridge Univ. Press, 2004.

\end{thebibliography}

\end{document}